\documentclass[useAMS,usenatbib]{mn2e}
\usepackage{subfigure}

\usepackage{graphicx}
\usepackage{times}%
\usepackage{natbib}

\newcommand{\gtrsim}{\ga}

\def\etal{et al.}

\def\aap{A\&A}
\def\apj{ApJ}

\def\mnras{MNRAS}

\def\aj{AJ}

\def\nat{Nature}

\def\apjs{ApJS}

\def\ie{{\it i.e.} }
\def\CR{{\tt CRASH} }
\def\CRLya{{\tt CRASH$\alpha\,$}}
\def\lya{Ly$\alpha\,$}

\title[CRASH$\alpha$: coupling continuum and line radiative transfer]
{CRASH$\alpha$: coupling continuum and line radiative transfer.}
\author[M. Pierleoni, A. Maselli and B. Ciardi]
{M. Pierleoni$^{1}$, A. Maselli$^{1}$ and B. Ciardi$^{1}$\\
$^1$ Max-Planck-Institut fuer Astrophysik,
Karl-Schwarzschild-Strasse 1, D-85748 Garching b. Muenchen, Germany\\
}

\begin{document}

\date{Accepted ???. Received ???; in original form ??? 2007}
\pagerange{\pageref{firstpage}--\pageref{lastpage}} \pubyear{2007}
\maketitle
\label{firstpage}

\begin{abstract}
In this paper we present \CRLya, the first radiative transfer code
for cosmological application that follows the parallel propagation of \lya and
ionizing photons. \CRLya is a version of the continuum radiative transfer
code \CR with a new algorithm to follow the propagation of \lya photons through
a gas configuration whose ionization structure is evolving.
The implementation introduces the time evolution for \lya photons (a feature
commonly neglected in line radiative transfer codes) and, to reduce the computational
time needed to follow each scattering, adopts a statistical approach to the \lya
treatment by making extensive use of pre-compiled tables. These tables
describe the physical characteristics of a photon escaping from a gas
cell where it was trapped by scattering as a function of the
gas temperature/density and of the incoming photon frequency.
With this statistical approach we experience a drastic increase  of
the computational speed and, at the same time, an excellent agreement with
the full \lya radiative transfer computations of the code MCLy$\alpha$.
We find that the emerging spectra keep memory of the ionization history which generates
a given ionization configuration of the gas and, to properly account for this effect,
a self-consistent joint evolution of line and ionizing continuum radiation as implemented
in \CRLya is necessary. A comparison between the results from our code and from
\lya scattering alone on a fixed HI density field shows that the extent
of the difference between the emerging spectra depends on the
particular configuration considered, but it can be substantial and can
thus affect the physical interpretation of the problem at hand. These
differences should furthermore be taken into account when computing
the impact of the \lya radiation on  {\it e.g.} the observability of
the 21~cm line from neutral hydrogen at epochs preceeding complete reionization.
\end{abstract}



\section{Introduction} \label{sect:intro}

The detection of \lya lines from local and distant objects has always
been of great importance in astrophysics. It has been extensively used
as indicator of redshift, as a measurement of the star formation
activity of galaxies and as a probe of their internal structure. 
In the last few years an increasing interest has been devoted to the
search of \lya emitters (LAEs) at high redshift, which
are  expected to be characterized by a strong \lya
emission (Partridge and Peebles 1968), but               
significantly attenuated by dust absorption. In fact, 
it has been necessary to wait for dedicated large programs of deep
narrow band searches like the Large Area Lyman Alpha (LALA) and the Subaru
Deep Field survey to detect a significant number of emission galaxies
at high redshift (e.g. Stern et al. 2005; Iye et al. 2006) and to get complete
spectroscopic samples of LAEs at redshift $z=4.5$, $z=5.7$ and $z =6.5$ 
(e.g. Hu et al. 1998, 2004; Rhoads \& Malhotra 2001;
Kodaira et al. 2003; Taniguchi et al. 2005; Kashikawa et al. 2006; 
Murayama et al. 2007; Dawson et al. 2007).
Strong lensing magnification has been necessary to move the detection frontiers even
further, with several
candidates currently observed up to  $z\simeq 10$ 
({\it e.g.} Pell\'o et al. 2007; Stark et al. 2007).

The intense activity reported above is explained by the great  
interest in using LAEs as cosmological probes: LAEs are in fact 
the objects with the highest known $z$ and can be used to
study large scale structures and galaxy formation in the high redshift
universe. Number counts, together with the statistics of line
shapes, are extremely powerful observables from which inferring 
important information about {\it e.g.} the
properties of the intergalactic medium (IGM) and the reionization
era (e.g. Hu 2002; Rhoads 2003; Stern 2005), the  photoionization processes
and UV photon production (e.g. Stark 2007), the tomography of neutral gas,
gas velocity field and star formation activity (e.g. Kodaira et al. 2003).
This predicting power relies on the fact that the \lya line is shaped
inside the galaxy interstellar medium and at high
redshift is also affected by the IGM opacity, which becomes non
negligible to \lya photons at $z\gtrsim 6$ (Fan et al. 2006).

Emission of \lya photons from high redshift sources has also an impact
on the detectability of 21~cm line from neutral hydrogen in the IGM. 
At high redshift, the Wouthuysen-Field effect (Wouthuysen 1952; Field
1958, 1959) is in fact extremely efficient in decoupling the spin
temperature of the gas, $T_s$, from the cosmic microwave background (CMB)
 temperature, $T_{\rm CMB}$, 
allowing the 21~cm signal to be visible either in absorption or in emission.
Fluctuations in the \lya flux, due both to inhomogeneous distribution
of the \lya radiation sources and to the scattering in the wings, 
can modify the expected signal (e.g. Barkana \& Loeb 2005; 
Chen \& Miralda-Escud\'e 2006; Chuzhoy \& Zheng 2007; 
Semelin, Combes \& Baek 2007), at least
as long as a strong \lya background is not established and the
radiation intensity reaches a saturation level (e.g. Ciardi \& Madau
2003; Ciardi \& Salvaterra 2007). For these reasons, it is important
to follow the propagation of \lya photons rather than assume a
homogeneous background as it is generally done.

Due to the resonant nature of the \lya line propagation, a self-consistent
and detailed treatment of the line radiation transfer is
required in order to model properly how \lya radiation 
affects the IGM, as well as to understand how different 
physical processes shape the spectral features of LAEs. 
As a consequence of the great interest in this field, several
semi-analytic and numerical studies of the \lya radiative transfer
have followed the first pioneering papers on the subject (Osterbrock
1962; Avery \& House 1968; Adams 1972; Harrington 1973; Neufeld 1990).
Analytic solutions have been derived only for few simple geometrical gas
configurations: static plane parallel slabs including dust
(Neufeld 1990), static uniform sphere (Dijkstra, Haiman \& Spaans 2006) and
uniform gas with pure Hubble flow around a steady \lya source (Loeb \&
Rybicki 1999).
Given the difficulties in the treatment of radiative transfer though,
also the numerical approaches developed so far, mostly based on Monte
Carlo techniques, have been in most cases 
specifically designed for particular physical configurations and 
problems: 1D dusty and optically thick media (Ahn, Lee \& Lee 2000, 2001); 
3D arbitrary distribution of dustless gas with arbitrary bulk velocity field
(Zheng \& Miralda-Escud\'e 2002); spherically symmetric collapsing gas
clouds (Dijkstra, Haiman \& Spaans 2006); \lya scattering off opaque,
dusty and moving clouds (Hansen \& Oh 2006); Hubble like expansion
flows of neutral gas (Loeb \& Rybichi 1999; Kobayashi \& Kamara 2004).
Other codes have been specifically designed for studying LAEs and \lya
pumping in a cosmological context (Gould \& Weinberg 1996; Cantalupo
et al. 2005; Tasitsiomi 2006; Semelin, Combes \& Baek 2007).
Verhamme, Schaerer \& Maselli (2006, VSM06) have developed
a general-purpose 3D \lya radiation transfer code applicable to dusty 
media with arbitrary geometries and velocity fields.

So far analytical, semi-analytical as well as
numerical studies perform the \lya radiative transfer as
a post-process calculation by assuming a fixed ionization structure of
the gas through which it propagates, 
while none of them has tackled the \lya radiative transfer problem by taking into
account the effect of an evolving ionization configuration.
Nevertheless, as we show in the following of this paper, this approximation
results to be a poor one for some applications of interest, in particular
for cosmological studies at high redshift, but also when modeling the 
\lya emission from young galaxies.

In this paper we present \CRLya, a new radiative transfer scheme
which, for the first time in the literature, follows simultaneously 
the propagation of \lya and ionizing radiation self-consistently. 
This allows us 
to investigate the effects of evolving ionization configurations
on the propagation of \lya radiation and on the shaping of the line
emerging from single objects.
The impact of an evolving ionization structure can in fact be
significant and needs to be taken into account: 
the large cross-section of \lya photons makes propagation dominated by
resonant scattering with HI atoms and the random-walk-like nature of
the process makes the characteristic time for \lya photon
propagation much larger than the one for ionizing radiation. 
If an ionizing continuum changes the ionization of the gas through
which the \lya photon is propagating, the amount of scattering
suffered by the line photons before escaping will depend on the
ionization history of the system.
In this case, a joint treatment of both line and continuum transfer is needed
to study the alterations in the \lya spectrum occurring during the
evolutionary stages of the ionized regions. 

The code presented in this paper is the first step in this
direction. \CRLya has been implemented as an extension of the 3D ray-tracing
radiative transfer code for ionizing  radiation \CR (Ciardi et al. 2001; Maselli, Ferrara \& Ciardi 2003;
Maselli \& Ferrara 2005; Maselli, Ciardi \& Kanekar 2008), by
developing a new independent algorithm which follows
the path of line photons in time and space. As described in details in
the following, this new algorithm makes extensive use of pre-compiled
tables which have been derived by using the line transfer code
MCLy$\alpha$ (VSM06) and allows to compute in an extremely efficient way the
path of line photons in arbitrary 3D gas distributions. 

The paper is structured as follows. Section 2 is dedicated to a brief
overview of \CR and MCLy$\alpha$, while in Section 3 we describe the new
method. Some validation tests are shown in Section 4. In the last Section
we present a summary of the paper.


\section{CRASH and MCLy$\alpha$} \label{sect:lum}
In this Section we give a brief description of the codes 
\CR and MCLy$\alpha$ for the sake of providing a proper background
for the description of \CRLya given in Section 3.
A more detailed description of the two codes is already in the 
literature: the details of \CR implementation are given mostly in 
Maselli, Ferrara \& Ciardi (2003), with updates on 
a new scheme for the background radiation field  given in 
Maselli \& Ferrara (2005) and on the latest version of the code
in Maselli, Ciardi \& Kanekar (2008).
MCLy$\alpha$ algorithm is fully described in VSM06. 
Note that some nomenclature has been changed for clarity.

\subsection{CRASH: continuum radiative transfer}
\label{crash3}

\CR is a 3D ray-tracing radiative transfer code based on 
Monte Carlo (MC) techniques that are used to sample the probability distribution
functions (PDFs) of several quantities involved in the calculation, {\it e.g.}
spectrum of the sources, emission direction, optical depth.
The MC approach and the code architecture assure a great
flexibility in the application to a wide range of astrophysical
problems and allow additional physics to be easily added with a minimum effort. 

The algorithm follows the propagation of the ionizing radiation
through an arbitrary H/He static density field and at the same time
computes the  variations in temperature and  ionization state of the
gas. Both multiple point sources, located arbitrarily in the box,
 and diffuse radiation ({\it e.g.} the
ultraviolet background or the radiation produced by H/He recombinations) can be accounted for. 
In this paper we neglect the treatment of any background radiation for
simplicity.

The energy emitted by point sources in ionizing radiation is discretized into
photon packets, beams of ionizing photons, emitted at regularly spaced time intervals.
More specifically, the total energy  radiated by a single source of luminosity
$L_s$, during the total simulation time, $t_{sim}$, is
$E_s=\int_0^{t_{sim}}L_s(t_s)dt_s$. For each source, $E_s$ is distributed
in $N_p$ photon packets, emitted at the source location at regularly spaced time intervals, $dt=t_{sim}/N_p$. 
The time resolution of a given run is thus fixed by $N_p$ and
the time evolution is marked by the packets emission: the {\it j}-th  packet is 
emitted at time $t_{em,c}^j=j \times dt$, with $j=0,...,(N_p-1)$.  
Thus, the total number of emissions of continuum photon packets is 
$N_{em,c}=N_p$.
In its latest version (Maselli, Ciardi \& Kanekar 2008), the code allows for
polychromatic packets whose content consists of photons distributed 
in various frequency bins which are populated
according to the spectral shape assigned to the source.

The emission direction of each photon packet is assigned by MC sampling 
the angular PDF characteristic of the source. 
The propagation of the packet through the given density
field is then followed and the impact of radiation-matter interaction on the gas
properties is computed on the fly.      
Each time the packet pierces a cell $i$, the cell optical depth for
ionizing continuum radiation, $\tau_c^i$, is estimated summing up the
contribution of the different absorbers ($\rm HI$, $\rm HeI$, $\rm HeII$).
As the probability for a single photon to be absorbed in such a cell is:
\begin{equation}
\label{tauPDF}
P(\tau_c^i)=1-e^{-\tau_c^i},
\end{equation}
the number of photons absorbed in the cell $i$ is the fraction
$P(\tau_c^i)$ of packet content when entering the
cell. In the polychromatic implementation, the same argument applies
to the number of photons contained in each single frequency bin.
The trajectory of the packet is followed until its photon content is
extinguished or, if continuum boundary conditions are not assumed,
until it exits the simulation volume.

The time evolution of the gas physical properties (ionization
fractions and temperature) is computed solving in each cell the appropriate
discretized differential equations each time the cell is crossed by a
packet. The reader is referred to Maselli, Ferrara \& Ciardi (2003) and
Maselli, Ciardi \& Kanekar (2008) for more details.

\subsection{MCLy$\alpha$: line radiative transfer}
\label{mclya}

MCLy$\alpha$ is a numerical scheme for Ly$\alpha$ line radiative
transfer, whose implementation is based on the basic structure of
\CR. 
MCLy$\alpha$
in fact uses the same MC sampling and ray-tracing techniques and it allows for arbitrary 3D  
hydrogen plus dust density distributions, as well as for arbitrary ionization, temperature and
velocity fields.

There are three physical processes, included in the code, which affect
the propagation of the line radiation: Ly$\alpha$ line scattering,
dust absorption and dust scattering.
For the sake of simplicity, in this paper we concentrate solely on the effect of
Ly$\alpha$ line scattering and we defer the treatment of the interaction
between radiation and dust to future work. 
Here we describe the basic structure of the algorithm in the absence of
dust. For a more complete and accurate description the reader is
referred to the original paper (VSM06).

Ly$\alpha$ is the strongest $\rm HI$ transition, 
for which the cross-section assumes large values at  
frequencies near the line center, $\nu_0=2.466 \times 10^{15}$ Hz.
It is convenient to introduce the frequency shift:
\begin{equation}
x=\frac{\nu-\nu_0}{\Delta\nu_D},
\end{equation}
where $\Delta\nu_D=(V_{th}/c) \nu_0$ corresponds to the Doppler frequency width and $V_{th}$ is the 
velocity dispersion of the Maxwellian distribution describing the thermal motions, {\it i.e.}
$V_{th}=\left( 2k_BT/m_H\right)^{1/2}=12.85 \; T_4^{1/2}$~km~s$^{-1}$,
with $T_4$ being the gas temperature in units of $10^4$~K. The other symbols have the usual meaning.
Here we neglect turbulent motions, but the option is available for their inclusion.

The  \lya  line radiation field is reproduced by emitting photons
from each source and by following their path through
the assigned gas distribution until they escape from the simulation box.
The location of interaction between the \lya photons and the gas is
determined by MC sampling the PDF for the line optical depth a photon
crosses before being scattered, $P(\tau_l)=1-{\rm e}^{-\tau_l}$. 
In other terms, the location of interaction
is determined as the cell at which the total optical depth from the 
emission location, $\tau_l=\sum_i \tau_l^i$ (where the sum extends over all
the cells crossed by the photon), becomes larger than
$\tau_{scatt}=-\ln(1-\xi)$, where $\xi$ is a random number extracted in the interval $[0:1[$.

The next step, after assessing the absorption location, is to determine
the photon frequency following a scattering with a hydrogen atom. To do
this, the code first converts the frequency of the photon 
from the external (observer) frame, $\nu_{obs}$, to the one comoving with
the fluid, $\nu_{com}$, performing a Lorentz transformation:
\begin{equation}
\nu_{com}=\nu_{obs}\left( 1-\frac{\textbf{k}_{in}\cdot \textbf{V}}{c}\right),
\end{equation}
where $\textbf{k}_{in}$ is the incoming photon direction and $\textbf{V}$ the bulk
velocity of H atoms. Due to the thermal motion of H atoms, scattering
in the fluid comoving frame is not perfectly coherent.
Within the comoving framework\footnote {
Here and in the following we omit the comoving suffix for the sake of
keeping an easily readable notation.} and neglecting the recoil
effect, partially coherent scattering can be described with a simple relation
between the incoming, $x_{in}$, and the outcoming, $x_{out}$, frequency (Dijkstra, Haiman \& Spaans 2006):
\begin{equation}
x_{out}=x_{in}-\frac{\textbf{V}_a\cdot\textbf{k}_{in}}{V_{th}}+\frac{\textbf{V}_a\cdot\textbf{k}_{out}}{V_{th}}.
\end{equation}
In the above equation $\textbf{V}_a$ is the atom velocity, while $\textbf{k}_{in}$ and $\textbf{k}_{out}$
are respectively the incoming and outcoming propagation direction.
The code can model both isotropic and dipolar angular redistribution;
in this paper we use only the isotropic redistribution and sample randomly the outcoming
propagation direction.

Once a new direction and frequency are assigned to the scattered
photon, a new random $\xi$ is extracted to determine the next
scattering location. 

This scheme is repeated until the photon escapes
the simulation volume.

\section{CRASH{\large $\alpha$}: combining continuum and line transfer}

In this Section we describe \CRLya, the first numerical
scheme which combines the treatment of continuum and line transfer
radiation.
As mentioned in the introduction, the algorithm has been developed as
an extension of \CR, which provides the treatment of the
ionizing radiation as described in the previous Section and references
therein.
The extension indeed consists in a new algorithm developed to follow
the propagation of \lya photons through a given gas configuration while
it is changed by ionizing radiation. In fact, although the continuum
photon propagation proceeds undisturbed by the \lya radiation field,
\lya radiative transfer is strongly affected by the change in the
ionization state of the gas.

In order to perform the coupling, it is necessary to introduce
the time evolution for \lya propagation, a feature commonly neglected
in line radiative transfer codes like MCLy$\alpha$.
This is a crucial aspect because, due to the resonant scattering
nature of \lya transfer in a neutral medium, \lya radiation can remain
trapped for a substantial fraction of the
simulation lifetime before being able to propagate away from its emission site, while the
propagation time of the ionization front can be much shorter. Thus, the change in the degree of
ionization affects the propagation of the \lya photons, while the
latter induces no back reaction on the gas. Note that, although \lya  photons, via
scattering,  can
transfer some of their energy to the gas and heat it, in typical situations 
the effect is negligible and thus such heating is not generally included in \lya radiative
transfer codes. We defer the investigation of this issue in more detail to future work.

To correctly model the simultaneous propagation of the two radiations a combined approach is needed.
This is a challenging task because of the very different nature of continuum and line
transfer, in terms of {\it e.g.} their path (straight line versus random walk) and
time-scales (see discussion above).
The above differences are reflected also in the numerical
implementation of line and continuum radiative transfer. For example,
while in the case of ionizing radiation the time needed for a photon
packet to travel a given distance does not depend sensibly on the
physical properties of the gas but only on the physical distance
crossed, the propagation of Ly$\alpha$ photons is very sensitive to
the ionization state of the gas and extreme configurations can be
faced, in which the \lya photons scatter for the entire simulation
time trapped in few cells without exiting the simulation volume.

As the ionizing radiation scheme has not been modified, it will not be 
discussed further and in the following we will focus on describing the
details of the line transfer part of the algorithm.

The Ly$\alpha$ radiation is discretized in a large number of photons
whose emission and propagation is dictated
by the time-scale attached to the ionizing radiation evolution. In
this way we are able to model the change in the
Ly$\alpha$ propagation due to the variations in the gas ionization
state. To correctly model the propagation of Ly$\alpha$ photons we
need to follow every single scattering.
As this would require a very large computational time, we use a
statistical approach to the Ly$\alpha$ treatment. We have compiled
1085 tables by running MCLy$\alpha$, in order to describe the physical
characteristics of a photon after a scattering depending on the
temperature and density of the gas and on the incoming photon frequency  (see Appendix).
The following part of this Section is dedicated to a description of the
various steps of the implementation.

\subsection{Emission of Ly$\alpha$ photons}
\label{lya_em}

Every Ly$\alpha$ emission is characterized by the generation of
$N_{\gamma,l}$ \lya line photons emitted at 
the same time, $t^i_{em,l}$. The parameter $N_{\gamma,l}$ is chosen to optimize the resolution
and the code performance. 
The code allows for two different methods for photon emission.
In the first method the emission is regularly spaced in time as in the continuum emission. If, as in
Section~\ref{crash3}, we define $N_{em,l}$ as the total number of emissions of line photons,           
in this case:

\begin{equation} 
\label{em_1}
t^i_{em,l}= i \times \frac{t_{sim}}{N_{em,l}},
\end{equation}
with $i=0,...,(N_{em,l}-1)$.\\
An alternative criterion for the emission follows the evolution of the
ionization structure.
In this case the emission time, $t^i_{em,l}$, is linked to the volume averaged H ionization
fraction, $\chi_{{\rm HII},em}$, 
and \lya photons are emitted at the time $t^i_{em,l}$ when:
\begin{equation} 
\label{em_2}
\chi^i_{{\rm HII},em}=i\times\Delta \chi_{\rm HII}.
\end{equation}
$\Delta \chi_{\rm HII}=1/N_{em,l}$ is the chosen HII fraction
variation in the gas and $i$ is an integer that covers values 
between $0$ and $N_{em,l}-1$.
While in the first formulation a constant \lya emission rate is
assured, in this case the emission rate is higher when the ionization
state of the gas changes faster. 
In order to reproduce a constant emissivity even in the second formulation, 
we assign a weight to each photon emitted at the $i$-th step: 
$w^i_{ph}=\left( t^i_{em,l}-t^{i-1}_{em,l}\right)/t_{sim}$.
When a \lya spectrum is built, each photon contributes according to its weight. This allows to
modulate the emission of \lya photons based on the change of the ionization degree (and thus
to better sample the effect of ionization on \lya scattering) and at the same time to 
have a constant \lya photon rate.

In the following tests the emission is assumed to be isotropic, but it
is always possible to account for an arbitrary angular PDF.

Every emitted Ly$\alpha$ photon $k$ ($k\in\left[1,N_{\gamma,l}\times
N_{em,l} \right] $) is described by its frequency in the comoving
frame $x_{in,k}$ (in this case we assume a monochromatic
spectrum with $x_{in,k}=0$, but a different spectrum can be used), 
position {\bf p}$_k$ (which coincides with the source location), 
direction of propagation {\bf k}$_{in,k}$, optical depth at which the scattering takes
place $\tau_{scatt,k}$ (as defined in Sec.~\ref{mclya})
 and a characteristic time $t_{ch,k}=t^i_{em,l}$ that is used to
evolve the photon along the simulation timeline (see next Section). At any step of the simulation
the $k$-th photon is always described by the quantities ($x_{in}$, {\bf p}, {\bf k}$_{in}$,
$\tau_{scatt}$, $t_{ch}$), where the index $k$ has been omitted for clarity. In the
following, we will always omit it. 

\subsection{Propagation of Ly$\alpha$ photons}

In Section \ref{crash3} we have seen how the physical time of the simulation is driven by the 
emission of packets of ionizing radiation discretized in time units, $dt$.
We are interested now to link the propagation of a Ly$\alpha$ photon to this timeline.

Let's assume that an ionizing photon packet has been emitted at $t_{em,c}^j$, that
the physical state of the gas has been evolved between $t_{em,c}^j$ and $t_{em,c}^{j+1}$,
and that a \lya photon is emitted at the same time $t^i_{em,l}=t_{em,c}^j$; 
then its characteristic time is assigned the value $t_{ch}=t^i_{em,l}$. 
The propagation of the \lya photon along the direction {\bf k} is 
followed between $t_{ch}$ and $t_{em,c}^{j+1}$, and the line optical depth encountered
along the path, $\tau_l$, is calculated as described in Section~\ref{mclya}. 
In each cell crossed by the photon we check if a scattering takes place, {\it i.e.}
if $\tau_l$ becomes larger than $\tau_{scatt}$.
If there is no scattering, we follow the propagation until $t_{em,c}^{j+1}$ and
at this point we store the photon's frequency $x_{in}$, the updated 
position ${\bf p}={\bf p}+(c~dt){\bf k}$,
and characteristic time $t_{ch}=t_{em,c}^{j+1}$. Propagation direction
{\bf k}$_{in}$ and optical depth for scattering $\tau_{scatt}$ remain 
unchanged.
These information will be used to follow the photon evolution in the next time unit.
We define this photon as ``active'', in the sense that it is not trapped by scattering
inside a cell but will resume its propagation in the next time unit.

Let's consider now the case in which the photon scatters during the time unit. 
Unlike MCLy$\alpha$, this code does not follow every scattering inside the cell, but determines the 
properties of the outcoming photon by interpolation of pre-compiled
tables (see Appendix A).
Given the gas temperature, $T_{cell}$, the line optical depth,
$\tau_{cell}$, of the 
cell where the scattering takes place, and the frequency, $x_{in}$, 
of the incoming photon, a linear interpolation of the tables is performed
to obtain the distribution of frequencies of the outcoming photon, $x_{out}$,
and of the time interval that the photon is expected to spend inside
the cell due to scattering, $t_{scatt}$.
From these distributions the code extracts the values for $x_{out}$
and $t_{scatt}$ that will be assigned to the photon.
This approach allows for a tremendous gain
in computational speed by adopting a statistical 
description of the scatterings that occur to the photon inside a cell, 
without following each one individually.  The characteristic time is updated as
$t_{ch}=t_{ch}+t_{scatt}$. If $t_{ch}>t_{em,c}^{j+1}$ the photon is put in a ``stand-by'' mode
and its propagation is resumed (with a new $\tau_{cell}$) only when the simulation time 
becomes larger than $t_{ch}$.

The procedure described above is repeated for (i) all the \lya photons
emitted at $t^i_{em,l}$, (ii) all the \lya photons ``active'' at
$t_{em,c}^j$ and (iii) all the \lya photons that exit the ``stand-by''
mode in this time unit. Then, a new ionizing photon packet is emitted
at $t_{em,c}^{j+1}$ and after it has been evolved up to
$t_{em,c}^{j+2}$ the \lya cycle starts again: all the ``active''
photons are evolved from $t_{em,c}^{j+1}$ to $t_{em,c}^{j+2}$; if
there are ``stand-by'' photons with $t_{em,c}^{j+1}<t_{ch}<t_{em,c}^{j+2}$
they are turned into ``active'' photons and evolved until
$t_{em,c}^{j+2}$; if new \lya photons are emitted in this time unit
they as well are evolved until $t_{em,c}^{j+2}$.

If the ionizing radiation crosses a cell in which a Ly$\alpha$ photon is trapped by scattering, the
change in the physical conditions of the cell should be taken into account, as this affect the
characteristics of the outcoming photon. As an example,        
let's assume that a Ly$\alpha$ photon scatters in a cell at $t_{s,0}$ and that the time at 
which it exits the ``stand-by'' mode is $t_{ch}=t_{s,2}$. If the ionizing radiation crosses 
that cell at a time $t_{s,1}$ such that $t_{s,0}<t_{s,1}<t_{s,2}$, the physical 
conditions in the cell change. 
To take into account the effect on $x_{out}$ and $t_{scatt}$, 
we recalculate them from the tables by using the values 
$T_{cell}(t_{s,1})$, $\tau_{cell} (t_{s,1})$ modified by the ionizing
radiation and $x_{in}(t_{s,0})$ as we do not have any information on
the frequency of  the Ly$\alpha$ photon at $t_{s,1}$.

\subsection{Spectrum of Ly$\alpha$ photons}
\label{spectrum}

When photons exit the simulation box, their frequencies are collected to calculate the outcoming time
integrated spectrum. As discussed in Section~\ref{lya_em}, each photon is counted according to
its weight. To show the probability distribution of the outcoming \lya radiation,
spectra are normalized to the sum of all weights. 
The profile of the final spectrum strongly depends on the choice of the integration time.
To build a spectrum we define an initial, $t_{out}^l$, and a final, $t_{out}^{l+1}$, time.
All the photons that escape from the box in the interval $]t_{out}^l;t_{out}^{l+1}]$ will
contribute to the spectrum of the source. At the next output, all the photons collected in the interval
$]t_{out}^{l+1};t_{out}^{l+2}]$ will be used to build the spectrum.
This procedure is followed until the end of the simulation.
As in the case of the emission described in Section~\ref{lya_em}, the spectra can be 
produced regularly spaced in time or linked to the evolution of the ionization structure.
Thus, the time of the outputs is regulated by equations~\ref{em_1} and~\ref{em_2}, where
$t_{em,l}^i$ and $N_{em,l}$ are replaced by $t_{out}^l$ and $N_{out}$, respectively.

Spectra can also be built by choosing a pre-determined line of sight.

\section{Results}
\label{tests}

In this Section we perform tests for the parallel propagation
of ionizing and \lya radiation, which show how the evolution of the
ionization structure alters the \lya spectra of the outcoming radiation.
All the tests have the same initial conditions, unless stated otherwise.
We use a simulation box of 30~pc on a side, divided in $128^3$ cells.
A monochromatic ionizing source, emitting photons with energy equal to 13.6 eV,
is located at the center of the box; the ionizing photon rate is $5 \times 10^{49}
{\rm s}^{-1}$.
The ionizing radiation is discretized in $N_p=10^7$ photon packets.
The same source emits also a \lya monochromatic radiation.
As we want to construct spectra at a fixed distance from the source, we distribute
the gas (H only, with density $n_{\rm H}=1$~cm$^{-3}$, $N_{\rm HI}\sim5\times10^{18}$~cm$^{-2}$, and temperature
$T=10^4$~K) in a sphere of radius $r_{sph}=15$~pc around the central source. Outside the
sphere the density is set to zero, so that no interaction between radiation and
gas takes place. The gas is initially neutral.
Every simulation is carried out for a physical time of $t_{sim}=10^5$~yr.
In our reference runs we have $N_{out}=50$ outputs and $N_{em,l}=100$
emissions of \lya photons, each with $N_{\gamma,l}=10^4$ photons. Both the
emissions and the outputs are dictated by the evolution of the ionization field.
We will discuss the effect of a different choice for $N_{em,l}$, $N_{\gamma, l}$ and
$N_{out}$ at the end of the Section.
In the following we present the results of our simulations for different choices of
the dynamical state of the gas.
The spectra shown are obtained integrating on all directions in order to achieve a
better resolution, given the set of chosen parameters. 

\subsection{Static sphere}

\begin{figure*}
   \centering
   \includegraphics[width=7.5cm,angle=-90]{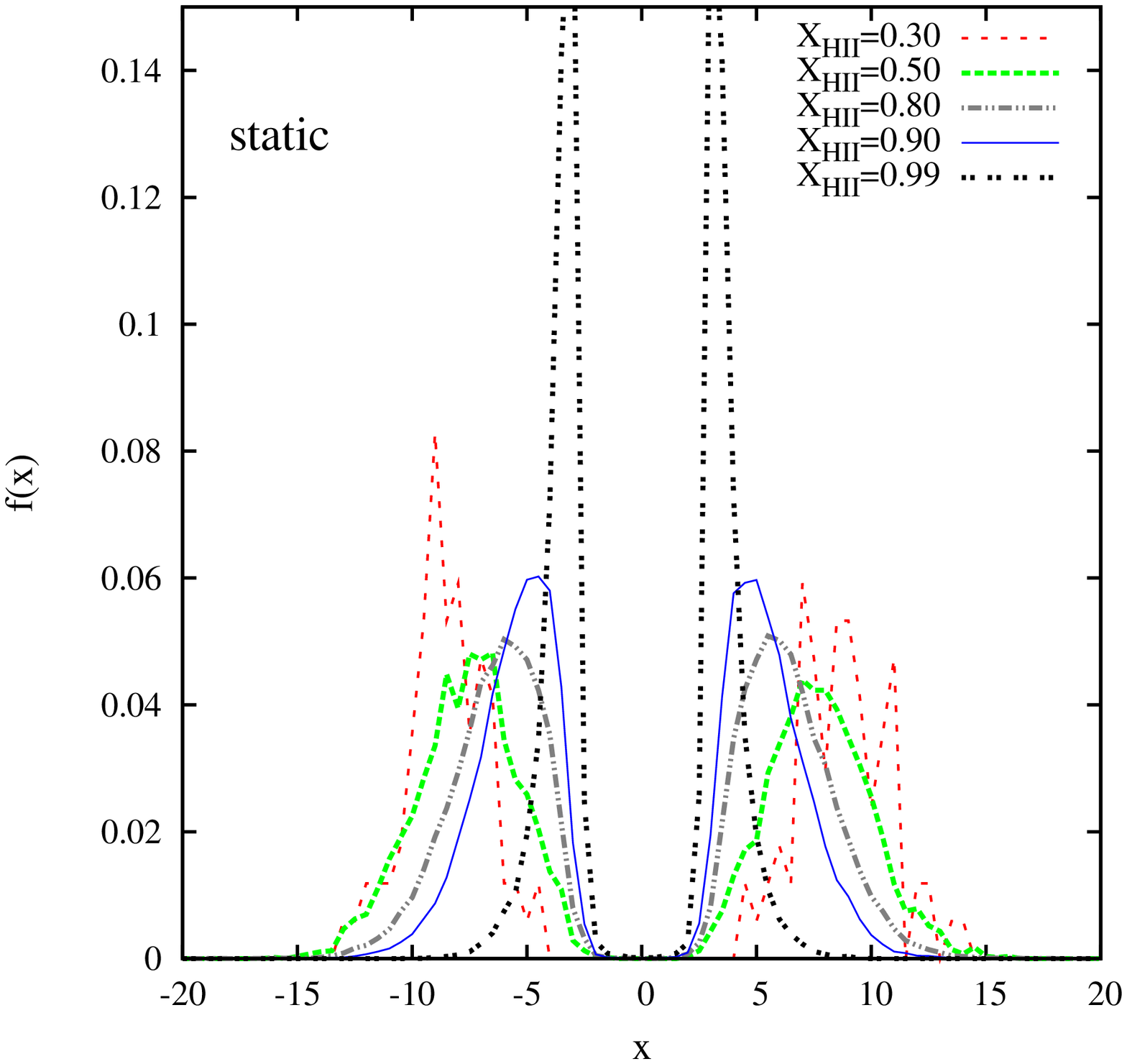}
   \includegraphics[width=7.5cm,angle=-90]{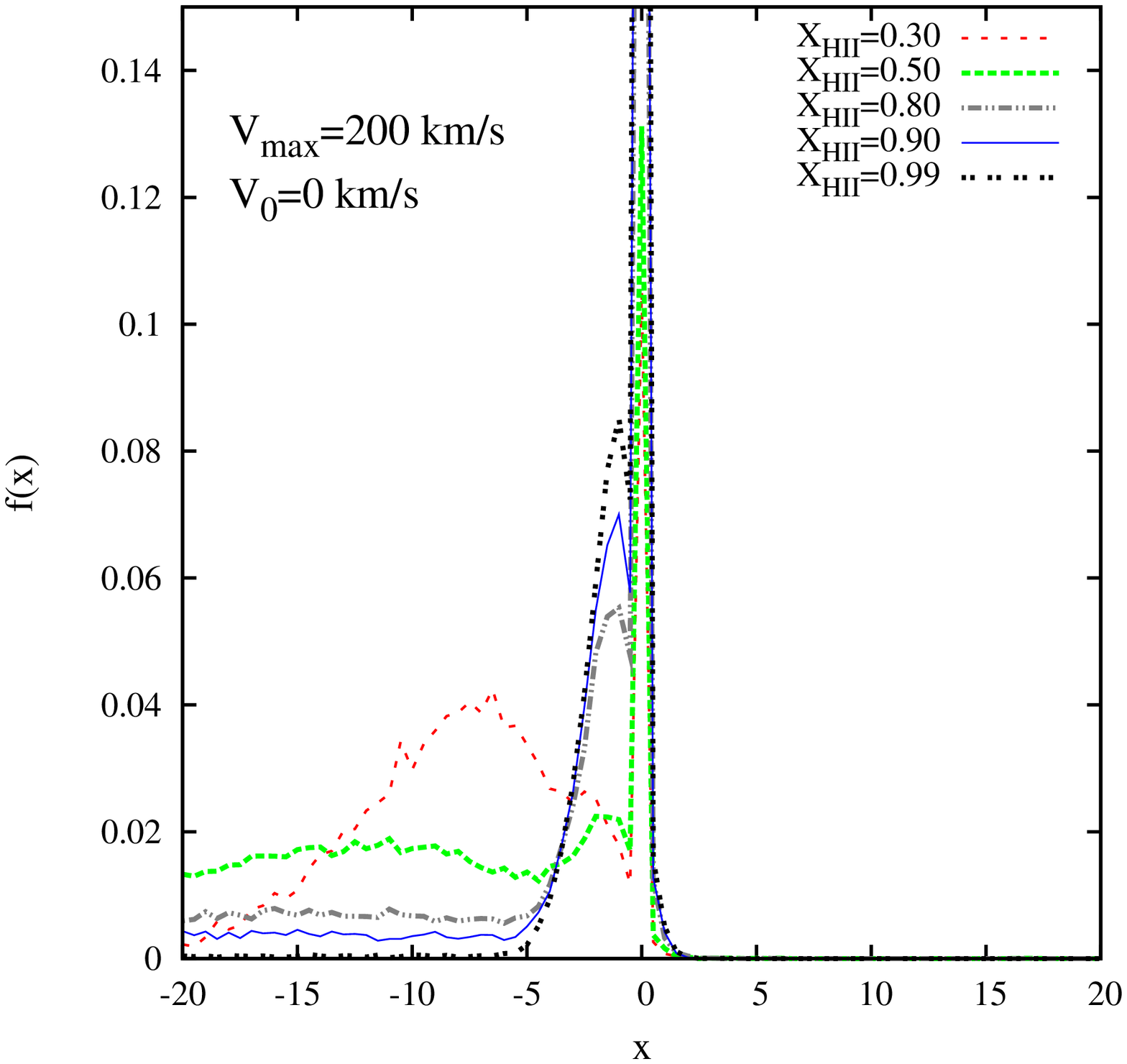}
   \includegraphics[width=7.5cm,angle=-90]{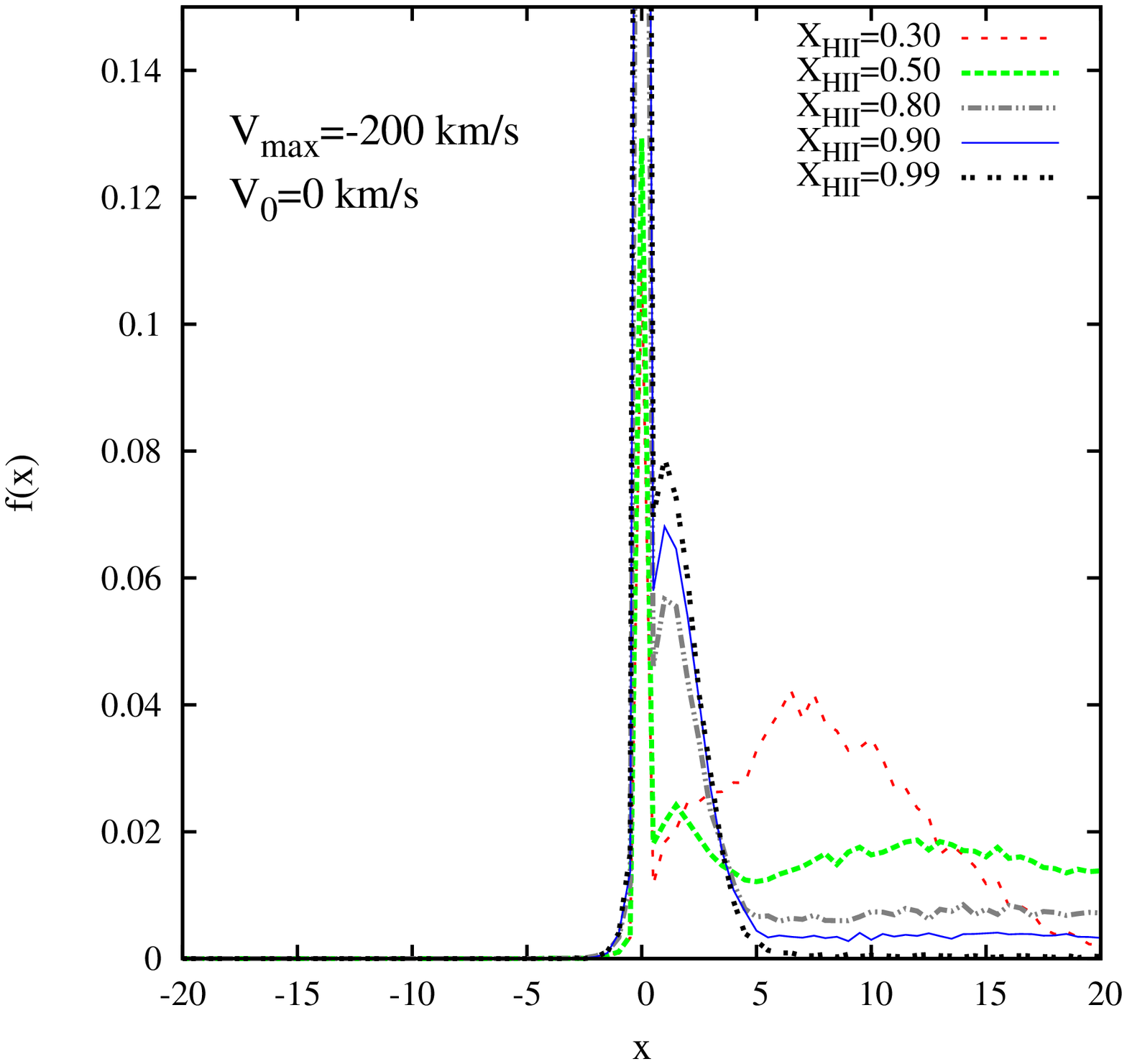}
   \includegraphics[width=7.5cm,angle=-90]{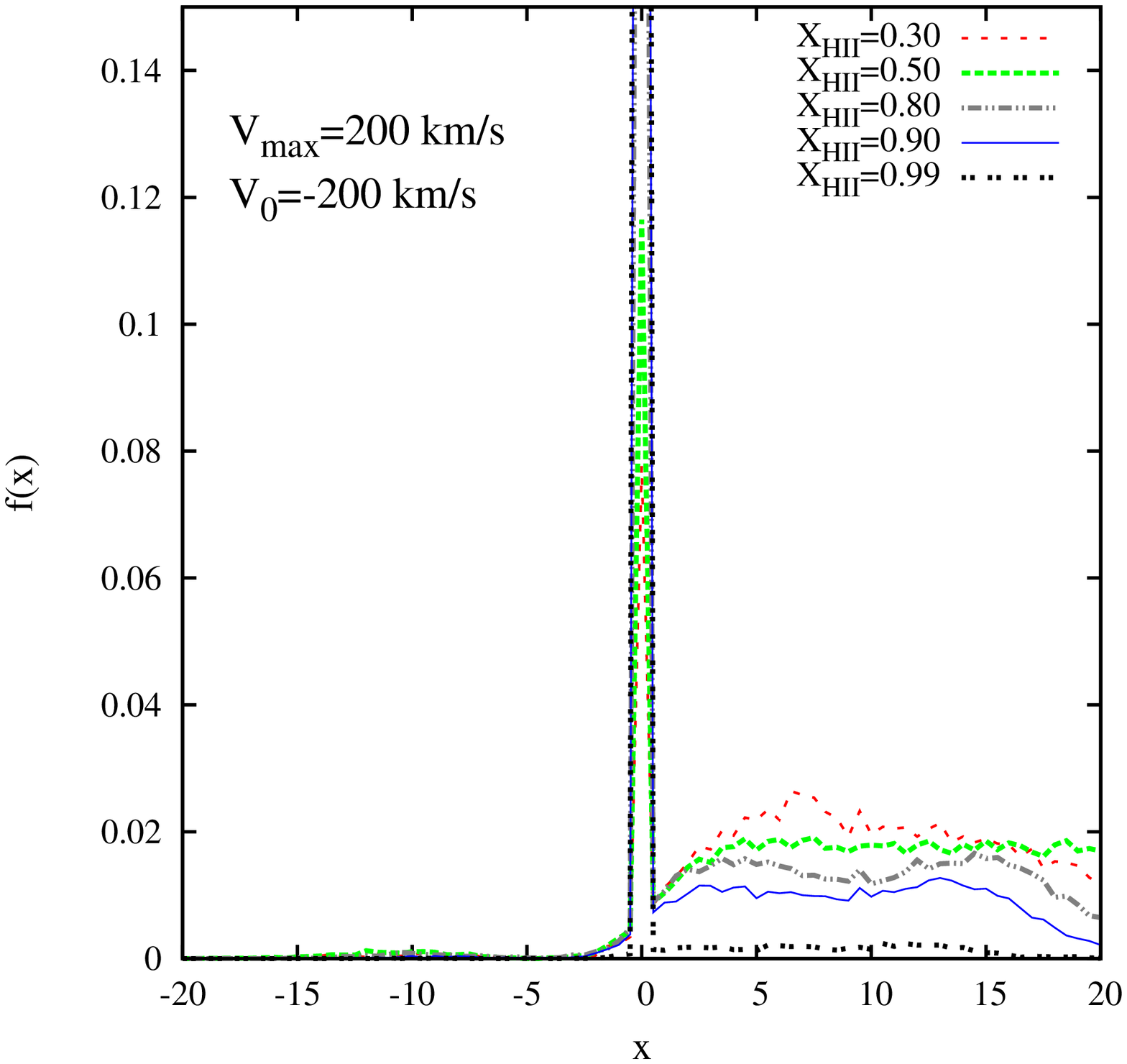}
   \includegraphics[width=7.5cm,angle=-90]{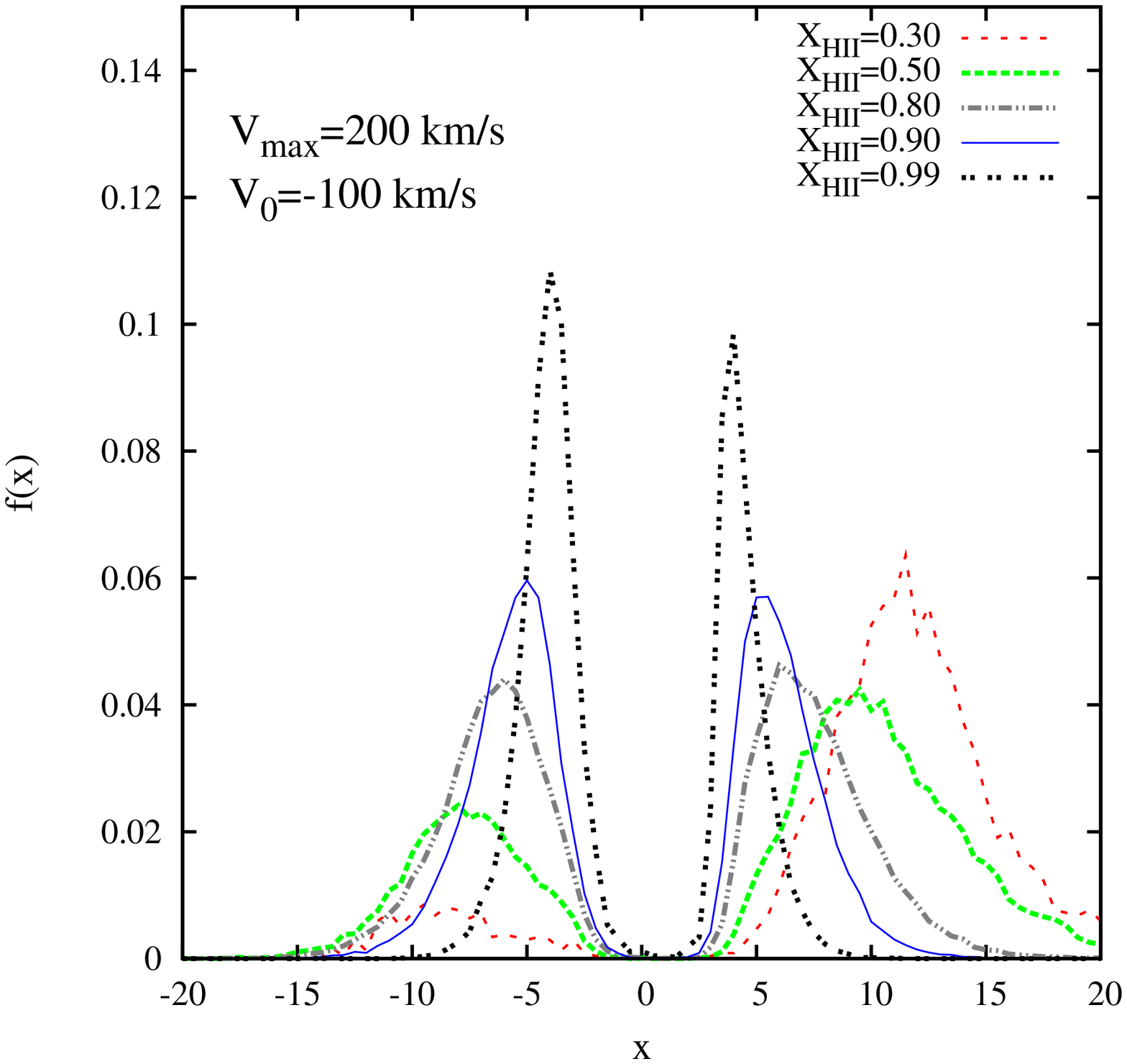}
   \includegraphics[width=7.5cm,angle=-90]{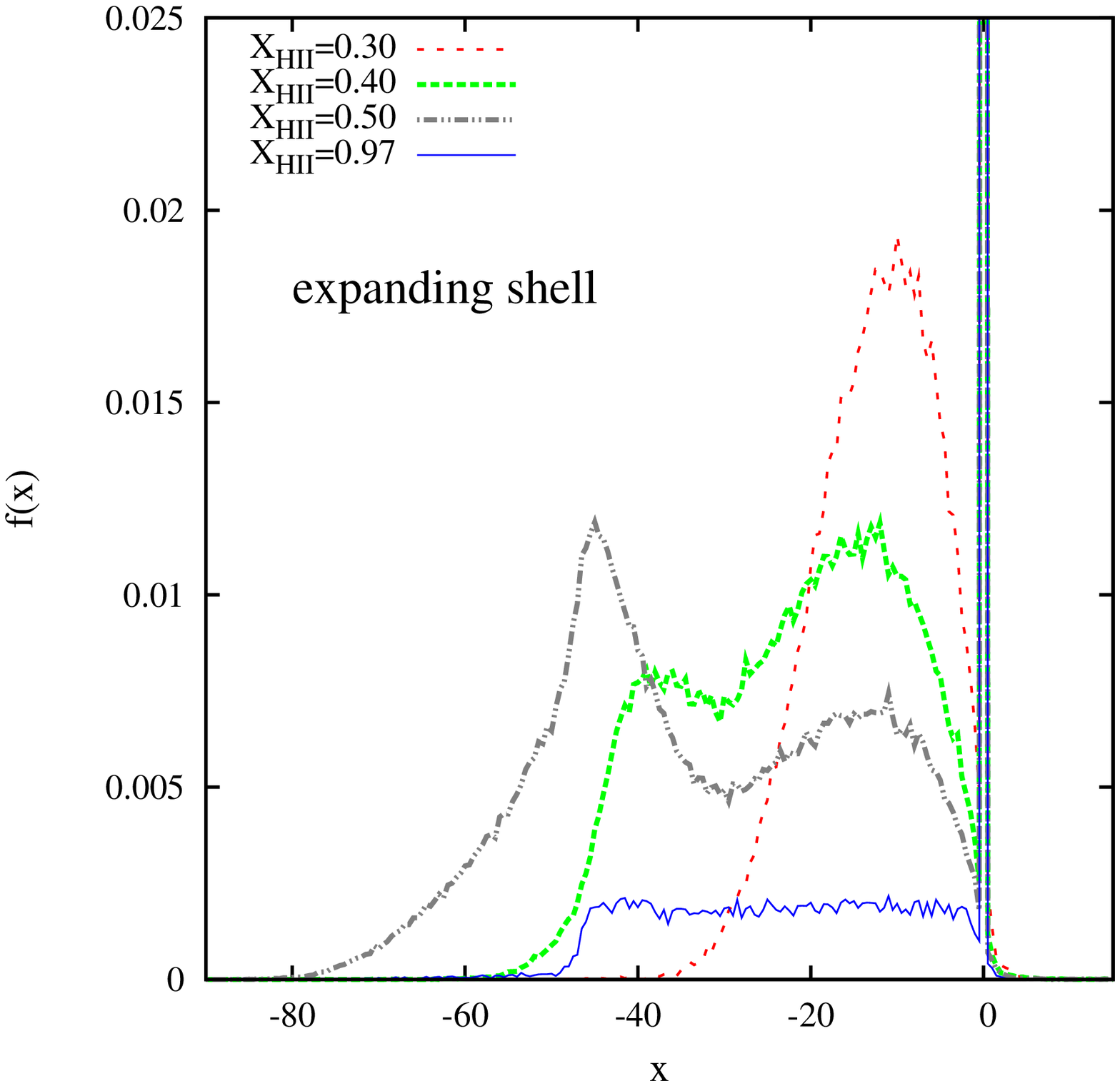}
  \caption{Spectra of \lya outcoming radiation at times corresponding to
ionization fractions $\chi_{\rm HII}= 0.3, 0.5, 0.8, 0.9, 0.99$
(with the exception of the bottom-right panel for which $\chi_{\rm HII}=0.3, 0.4, 0.5, 0.97$).
The dynamic condition of the gas is the following: static gas (upper-left panel),
homogeneous spherical cloud expanding and collapsing with velocity increasing
with distance from the source (upper-right and center-left panels), homogeneous spherical
cloud collapsing with velocity decreasing with distance from the source (center-right and bottom-left
panels), shell expanding at constant velocity (bottom-right panel). }
   \label{test_fig}

\end{figure*}

In this first test, the gas has no bulk velocity with respect to the
central source. The top-left panel in Figure~\ref{test_fig} shows the
spectra emerging from this configuration at times corresponding to
volume averaged
ionization fractions $\chi_{\rm HII}= 0.3, 0.5, 0.8, 0.9, 0.99$, which
could be regarded as spectra of the source observed at different times
elapsed since the source switches on.

The spectra shown here and in the rest of the paper
have been built as described in Section~3.3., {\it i.e.} collecting the \lya
photons escaping the system when $\chi_{\rm HII}$ falls in an interval
$\Delta\chi_{\rm HII}=0.02$ centered on the $\chi_{\rm HII}$ value
selected for the output.
The curve corresponding to $\chi_{\rm HII}=0.99$ is instead a
collection of the escaping photons  which starts
when $\chi_{\rm HII}=0.98$ and ends at $t_{sim}$.
For the combination of parameters chosen for the tests, at
$\chi_{\rm HII}<0.3$ the number of escaped \lya photons is not sufficient to build
a spectrum and also for $\chi_{\rm HII}=0.3$ the outcoming spectrum is very noisy.
As expected, the spectra exhibit the two symmetric peaks characteristic
of this configuration, although a direct, quantitative comparison with previous
works ({\it i.e.} Dijkstra, Haiman \& Spaans 2005) is not
possible, as none included the effect of ionizing radiation.
As the ionization increases, the peaks move towards $x=0$ because \lya photons encounter
less and less HI atoms along their path. At the same
time, the width of the peaks become smaller.
In this scenario we do not see a spectrum peaked at $x=0$ because the gas never gets
completely ionized and, for a static configuration, also a little fraction of neutral gas
far from the location of emission has a non negligible optical depth
for photons in the line center.

\subsection{Expanding and collapsing sphere}

In a more interesting case we simulate a homogeneous spherical cloud that collapses or
expands. In these tests we sample a velocity field, in the gas sphere,
described by $V(r)= V_{max} r/r_{sph} + V_0$.

Initially we choose $V_{max}=\pm 200$~km~s$^{-1}$ and $V_0=0$~km~s$^{-1}$.
The resulting spectra extracted at the same ionization fractions as for the static case
are shown in Figure~\ref{test_fig} with a positive and a negative
value for $V_{max}$ (top-right and center-left panels respectively).
As expected, the plots show specular \lya spectra, due to the opposite direction of the bulk motion.
When the gas is expanding the outcoming radiation is on
the red part of the spectrum, while it lays on the blue side when we consider
negative values for the velocity. This happens because the photons are
seen Doppler shifted according to the velocity of the atoms.
This means that, if an atom has a positive velocity, in the
atom rest frame a blue photon becomes a line center photon and is easily
blocked by the higher optical depth (compared to the optical depth in the wings).
Thus, a photon can escape only if it is shifted by scattering to the red side of the line.
Differently from the static case, here it is possible to have
\lya radiation at the central frequency $x=0$ also when ionization is not
complete. This is a consequence of the fact that, due to the Doppler
effect, the line center photons are seen in the atom rest frame as red (expanding gas)
or blue (collapsing gas) photons, \ie in the wing, and thus encounter a lower optical depth. 
In addition, the higher is the absolute value of the
velocity the bigger is the shift, so when the continuum radiation ionizes
the regions closer to the source (which have lower velocity and as a consequence
a higher contribution to the opacity), the optical depth at the center decreases significantly
because the external neutral layers of gas (with a higher velocity) give only a minor
contribution.
As a result, as ionization proceeds, we start seeing an increasing emission at the central 
frequency. More specifically, the spectrum corresponding to $\chi_{\rm HII}=0.3$ shows
radiation at $x=0$ and a residual in the red (blue) part of the
spectrum for positive (negative) velocities. As ionization proceeds, the
residuals become less pronounced and move towards the center, while the central radiation becomes stronger.
In the last spectrum, when the gas is 99\% ionized, there is still a
residual because of the remaining neutral hydrogen fraction in the
most distant regions of the gas sphere,
where the photo-ionization rate is suppressed by geometrical
dilution and by the residual inner opacity.

In the third case we consider a gas sphere collapsing with increasing velocity towards the center.
The center-right panel of Figure~\ref{test_fig} shows \lya profiles
generated with a bulk motion characterized
by $V_{max}=200$~km~s$^{-1}$ and $ V_0=-200$~km~s$^{-1}$.
These spectra are very similar to the ones obtained in the previous case, but as the
absolute value of the velocities increases towards the center, the residual blue part
of the spectra is more spread. Note that also in this case the residual is reduced as
ionization proceeds and the contribution to the opacity from the inner layers of gas
is suppressed.

In the last case (Fig. \ref{test_fig}, bottom-left) we consider
a gas which is collapsing near the source while the outer
shells are expanding, with $V_{max}=200$~km~s$^{-1}$ and $ V_0=-100$~km~s$^{-1}$.
The first \lya spectrum (at $\chi_{HII}=0.3$) is dominated by photons in the
blue part and just a residual is present on the red side. In fact, blue photons
have a larger probability to escape because the ionization front has not yet 
propagated far enough to suppress the contribution to the \lya gas opacity 
from the gas collapsing towards the source. 
As the front proceeds ionizing the neutral hydrogen with negative velocities
(spectra at increasing $\chi_{\rm HII}$), the red part of the spectrum becomes stronger 
while the blue part is suppressed, until, in the configuration with $\chi_{HII}=0.99$,
it is smaller than the red one.
As in the other tests, the increment in the ionization degree reduces the number of
scatterings, with the consequence of moving 
the two peaks towards the center, increasing their height and reducing their width.

\subsection{Expanding shell}
\label{shell}

In this Section we examine the case of a \lya source surrounded by 
an expanding shell, which has been extensively studied also by other authors
(Ahn, Lee \& Lee 2004; Hansen \& Oh 2006; VSM06).
Here, we are interested in the effects introduced on the \lya spectrum by an ionizing source 
inducing a time evolution of the neutral gas in the shell.
To simulate this configuration we have chosen a homogeneous density $n_{\rm H}=15$~cm$^{-3}$,
temperature $T=10^4$~K, and radial velocity $V=300$~km~s$^{-1}$.
All the gas is distributed within a shell of thickness 4~pc located at a distance of 10~pc 
from the source, while no gas is present outside the shell.
The corresponding column density is $N_{\rm HI}\sim2\times10^{20}$~cm$^{-2}$.
To show the \lya spectrum time evolution, we choose to plot the profiles 
corresponding to ionization fractions in the shell of
$\chi_{\rm HII}= 0.30, 0.40, 0.50, 0.97$ (Fig.~\ref{test_fig},
bottom-right panel).
For a better understanding of the spectral features, it is useful
to discuss the possible different paths for an outcoming photon.
Following VSM06 (see their Fig.~12), we divide the outcoming photons in three different
groups, depending on their scattering history:
\begin{itemize}
\item Backscattered photons: photons that, after scattering in the shell, travel inward
across the empty space before crossing again the shell. As these photons
undergo multiple scatterings with the gas, they can escape once they
are shifted on the red side of the line where the optical depth of the expanding
shell is smaller.
\item Diffused photons: all the photons which are diffused in the shell until
they escape without backscattering. We expect these photons to
contribute to a red bump in the spectra whose shift from the line
center and intensity will depend on the neutral gas density and on the
shell velocity. Typically the frequency shift will be smaller than for
backscattered photons as the number of scatters before escape is on average lower.
\item Directly escaped photons: photons that have no interaction with the gas and keep their initial 
frequency; in our case this group of photons will produce a peak at $x=0$.
\end{itemize}
It is important to underline that every group has a different 
characteristic time for escaping. In fact, directly escaped photons travel the shortest path.
On the other hand, diffused photons scatter in a volume smaller than the backscattered 
photons and thus escape faster;
therefore the red bump associated to the diffused photons will
typically appear before 
the feature produced by the backscattered photons.

Keeping in mind all the possible paths for \lya photons, let us analyze the features
in the spectra shown in the Figure~\ref{test_fig} (bottom-right panel).
The first profile ($\chi_{\rm HII}= 0.3$) exhibits a peak on the red side of the central emission,
due to the diffused photons, that, as already mentioned, escape faster than 
backscattered photons and can already been seen in the initial stages of the shell ionization.
Directly escaped photons are present as well and their abundance increases with time.
In the profile corresponding to an ionization degree of $\chi_{\rm HII}= 0.4$ we
can clearly see that a secondary bump is forming
at lower frequencies. At this stage of the evolution, the backscattering photons
are starting to escape from the shell with a frequency
that is more shifted respect to the other photons, as explained above.                                    
When the ionization degree is $\chi_{\rm HII}= 0.5$, the backscattering
bump is visible and dominant on the red peak due
to diffused photons.
In the profile corresponding to $\chi_{\rm HII}=0.97$ the ionization front has 
suppressed most of the neutral
gas and only a negligible fraction of \lya radiation interacts with the
residual gas in the shell; the result is a small fraction of photons
shifted on the spectrum's red side.

\subsection{Effect of ionizing radiation}

In the previous tests we have discussed how our time dependent treatment
of the \lya radiation allows to correctly establish the appearance at
different times of spectral features which are usually integrated in
the emergent spectra predicted with time independent formulations.
Here we investigate further on the importance of the joint propagation of \lya and
continuum radiation, by comparing results from two
different approaches to simulate the \lya spectra.
The first one, widely used in the literature, performs a \lya radiative
transfer on a gas configuration given as initial condition, which can be {\it e.g.}
a constant density field or a snapshot of a numerical simulation.
In this case, the gas configuration is kept constant throughout the entire
\lya radiative transfer and the \lya spectra are built once all the \lya photons
have escaped the simulation volume.
The other approach is the one described in this paper, \ie starting from an
initial gas configuration, the parallel propagation of continuum and line photons
is followed and the \lya spectra can be built at different times
taking into full account the changes in the \lya propagation due to the variations
in the neutral gas distribution.

\begin{figure}
   \centering
   \includegraphics[width=7.5cm,angle=-90]{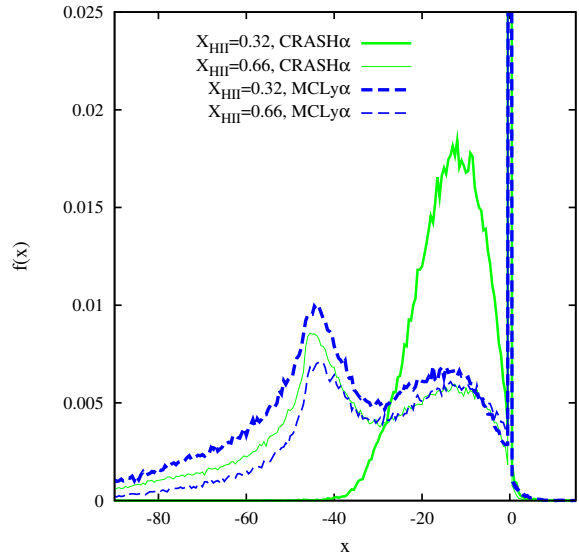}
  \caption{Comparison of \lya spectrum obtained with MCLy$\alpha$ and \CRLya
applied to the same gas distribution of an expanding shell.
For the application of MCLy$\alpha$ a fixed gas configuration is used,
while in our approach all the evolutionary stages are taken into account
(see text for details).}
\label{shell_McLya_fig}
\end{figure}

To show the impact of the two different approaches on the outcoming \lya spectra,
we consider the same configuration described in Section~\ref{shell},
\ie an initial neutral expanding shell which is ionized by a central
source emitting also \lya photons.
We compare the spectra obtained with \CRLya at the times when the
gas configuration is characterized by a mean ionization
fraction inside the shell of $\chi_{\rm HII}=0.32$ and~0.66, to those
obtained running MC\lya on the same gas configurations.
While with MC\lya the spectra are built by integrating over the
\lya photons once they have all escaped the fixed HI distribution,
the  \CRLya algorithm allows to account for the impact
on the emergent spectra of the ionization history which led to those
configurations.
The results are shown in Figure~\ref{shell_McLya_fig}.\\
A substantial difference is clearly visible in the spectra
corresponding to $\chi_{\rm HII}= 0.32$.
The \lya spectrum simulated by MCLy$\alpha$ is characterized by two
bumps associated with the diffused and backscattered photons.
A very different profile is obtained with \CRLya, where no
backscattered photon has escaped at this time and only the single red peak
corresponding to the diffused photons is present. As already underlined
in Section~\ref{shell} the absence of
backscattered photons is due to their larger escaping time.
The profiles corresponding to $\chi_{\rm HII}= 0.66$ are much more similar,
because at this stage a significant fraction of the backscattered
photons had enough time to escape the shell.
Nevertheless, there is still a difference in the amplitude of the peak
at $x=0$ and of the bumps from the backscattered photons, which
are also slightly more shifted. This difference is due to the
memory of \lya photons emitted in the previous stages of the source
activity, when $\chi_{\rm HII}< 0.6$.
In the MCLy$\alpha$ treatment all the emitted \lya photons see the same
mean shell opacity and the probability to have a direct
escape is significantly higher then in the \CRLya run, in which the \lya photons
see on average a larger shell opacity.
As a consequence, the fraction of photons with $x=0$ in \CRLya is smaller and, at
the same time, the photons that have remained trapped for a longer time exhibit
a larger amplitude of the spectra and a larger shift to the red side of the
central frequency.
The time elapsed between the two spectra considered above is only
about 100~yr. The time interval in which deviations from the
``instantaneous picture'' are significant is thus too small
to allow observations to capture these stages. However, the example is
useful to illustrate the relevant features and advantages of our
approach. Furthermore the deviations found could affect
the gas state, {\it e.g.} its spin temperature, independently from their 
observational detectability in the spectra.    

\begin{figure*}
   \centering
   \includegraphics[width=7.5cm,angle=-90]{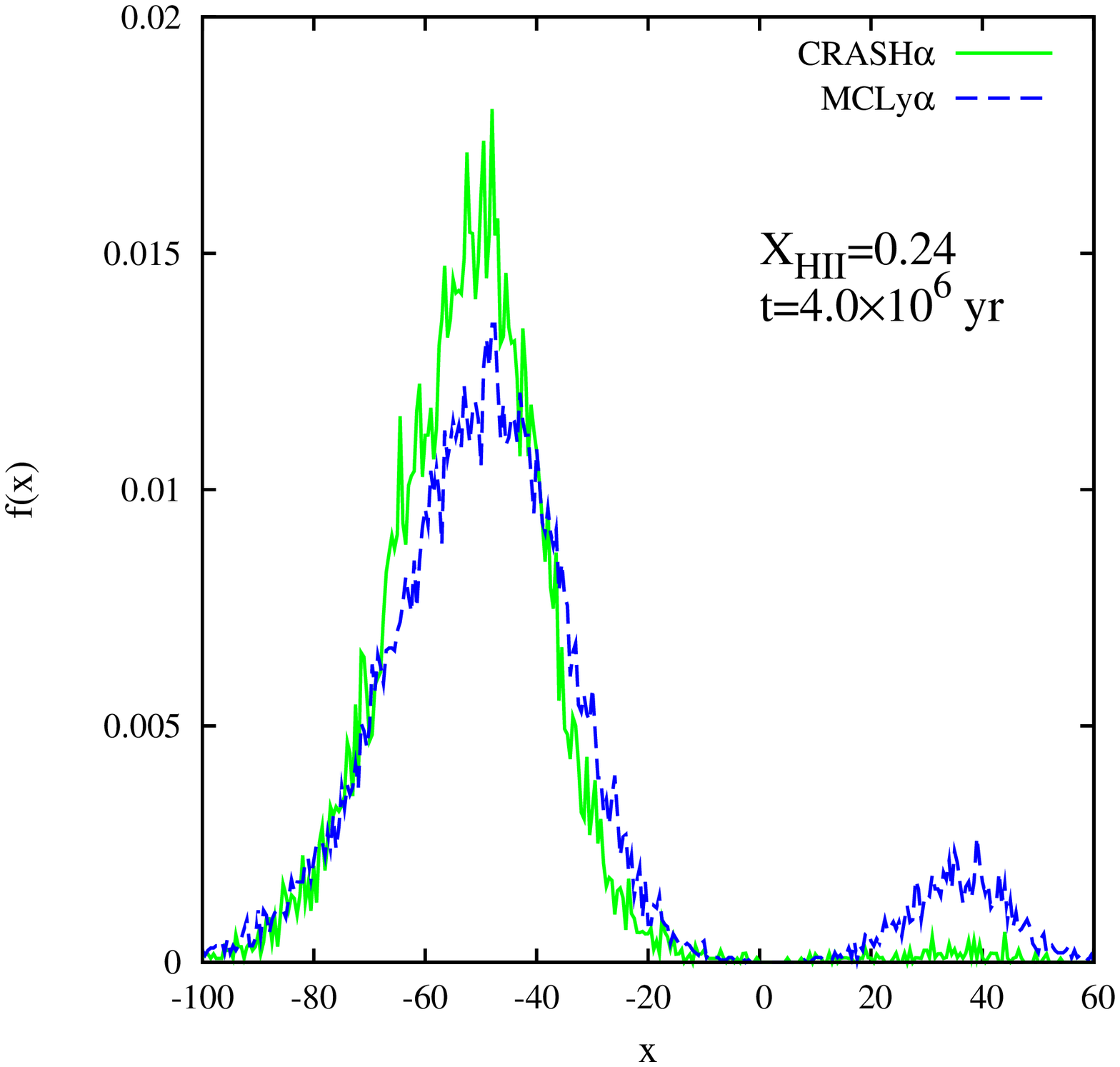}
   \includegraphics[width=7.5cm,angle=-90]{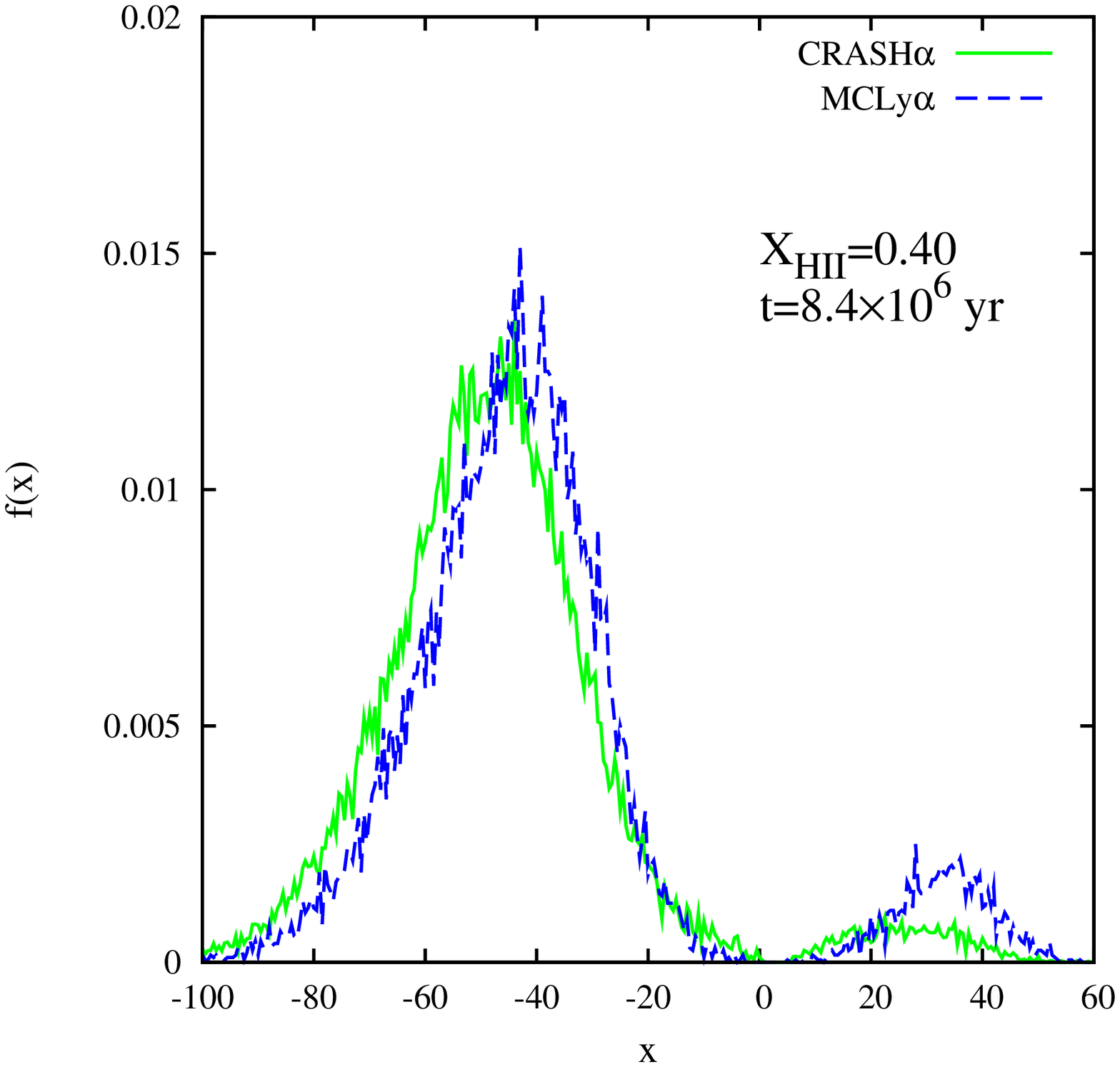}
  \caption{Comparison of \lya spectrum obtained with MCLy$\alpha$ and \CRLya
applied to the same gas distribution of an expanding sphere.
The left (right) panel shows \lya profiles at $t=4.0 \times 10^{6}$~yr
($t=8.4 \times 10^{6}$~yr). See text for more details.}
   \label{large_mclya}

\end{figure*}

A similar test has been performed to infer the observability of the
deviations in the spectra on larger scales. In this case
we use a simulation box of 200~kpc on a side and 
a photo-ionization rate of the central source $\dot{N}_\gamma=10^{54} {\rm s}^{-1}$.
In this test the gas is distributed in a sphere of radius
$r_{sph}=70$~kpc, with density $n_{\rm H}=0.01$~cm$^{-3}$
(corresponding to $N_{\rm HI}\sim2\times10^{21}$~cm$^{-2}$), while
we keep the temperature $T=10^4$~K; we also assume a velocity field
corresponding to a Hubble expansion with $H=$790~km~s$^{-1}$~Mpc$^{-1}$.
The simulation is carried out for a physical time of $t_{sim}=10^8$~yr.
In this case (Fig.~\ref{large_mclya}) the first spectrum is captured
at the time $t=4.0 \times 10^{6}$~yr (corresponding to 
a mean ionization fraction of $\chi_{\rm HII}= 0.24$) and the second
at the time $t=8.3 \times 10^{6}$~yr ($\chi_{\rm HII}= 0.40$).
While the MCLy$\alpha$ profile in the left panel of
Figure~\ref{large_mclya}, is characterized by a large red bump and a small blue one,
with the \CRLya approach we find a lower fraction of photons escaping
with blue frequencies and a larger red bump which is slightly shifted
on redder frequencies. This effect results from keeping memory of the
ionization history, since we take into account that before reaching
the observed gas configuration most of the \lya photons
have been trapped on the boundary of the growing ionized region.
The integrated optical depth along the full path traveled before
escaping is therefore larger respect to the one computed in the
MCLy$\alpha$ approach.
This effect is also visible in the right panel of
Figure~\ref{large_mclya}, where a larger red shift is present. 
In this case a larger fraction of trapped photons are now free to
escape and the shift is more evident.

The major result of these tests is that the emerging spectra keep memory
of the ionization history which generates a given observed
configuration and, to properly account for this effect,
the self-consistent joint evolution of line and ionizing continuum
radiation followed by our scheme is necessary.
The extent of the difference between the two methods depends on the
particular case considered, but it can be substantial and can thus affect the
physical interpretation of the problems at hand.
In a forthcoming study we will investigate in more detail which are the
objects/configurations for which the ionization effects are expected
to be relevant in shaping the observed \lya spectrum.

In addition, the time evolution that builds up the \lya radiation field can
be important when calculating the impact of such radiation on gas
properties like the spin temperature, which is relevant for the prediction
of the observability of 21~cm emission from neutral hydrogen 
at high redshift. We plan to include the computation of these effects
in a forthcoming extension of the code.   

\subsection{Dependence on input parameters}

\begin{figure*}
   \centering
   \includegraphics[width=7.5cm,angle=-90]{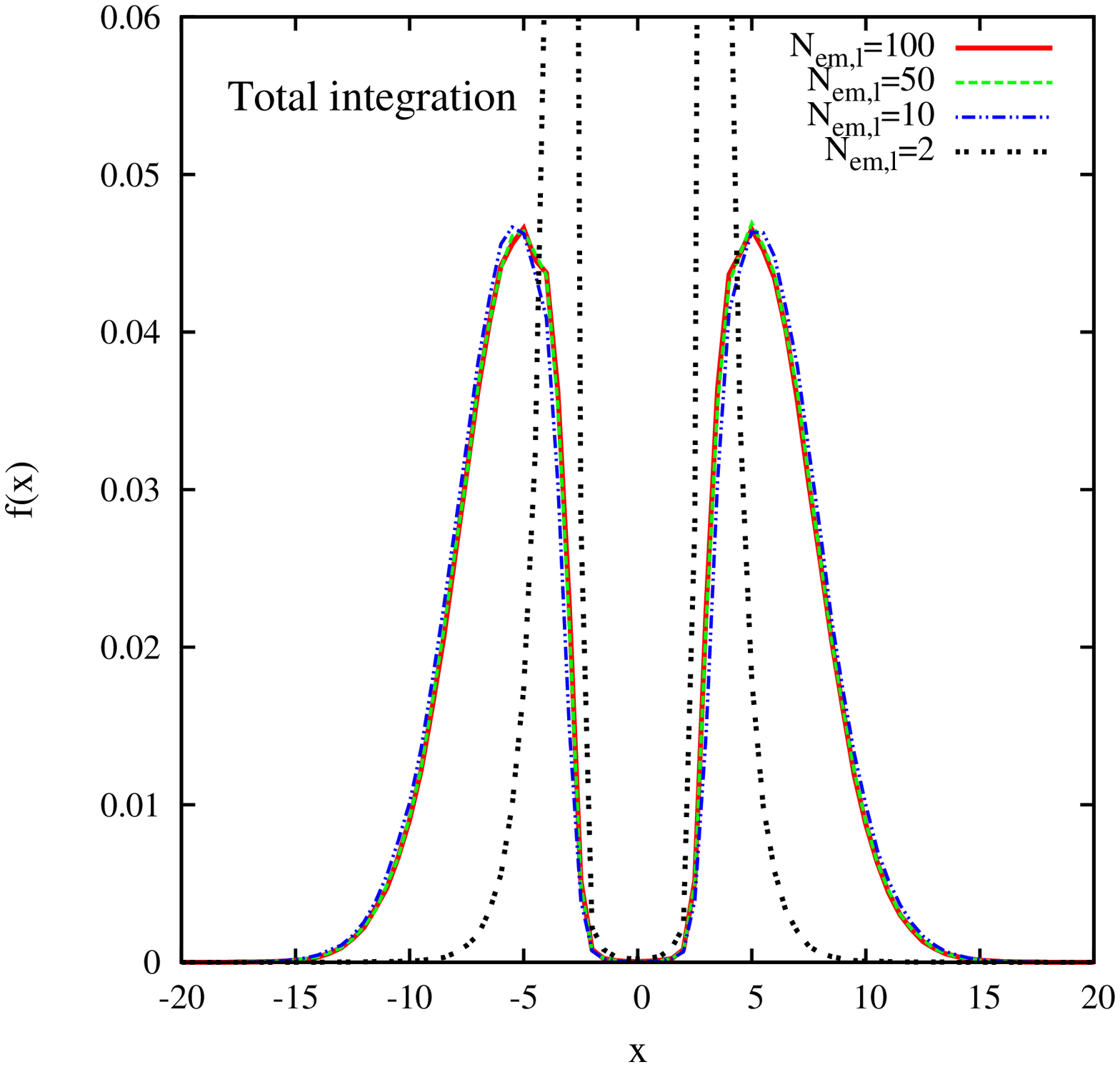}
   \includegraphics[width=7.5cm,angle=-90]{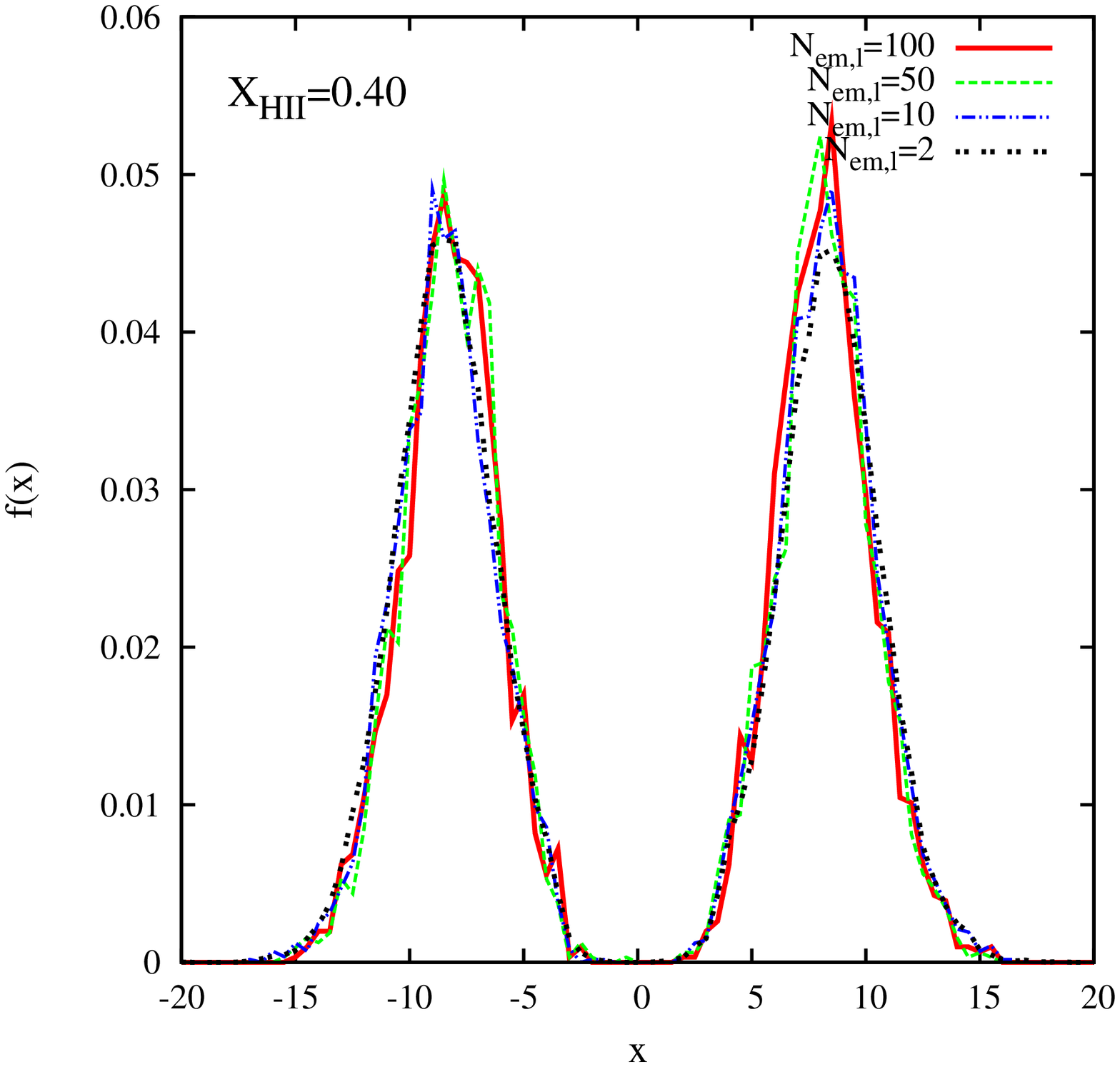}
   \includegraphics[width=7.5cm,angle=-90]{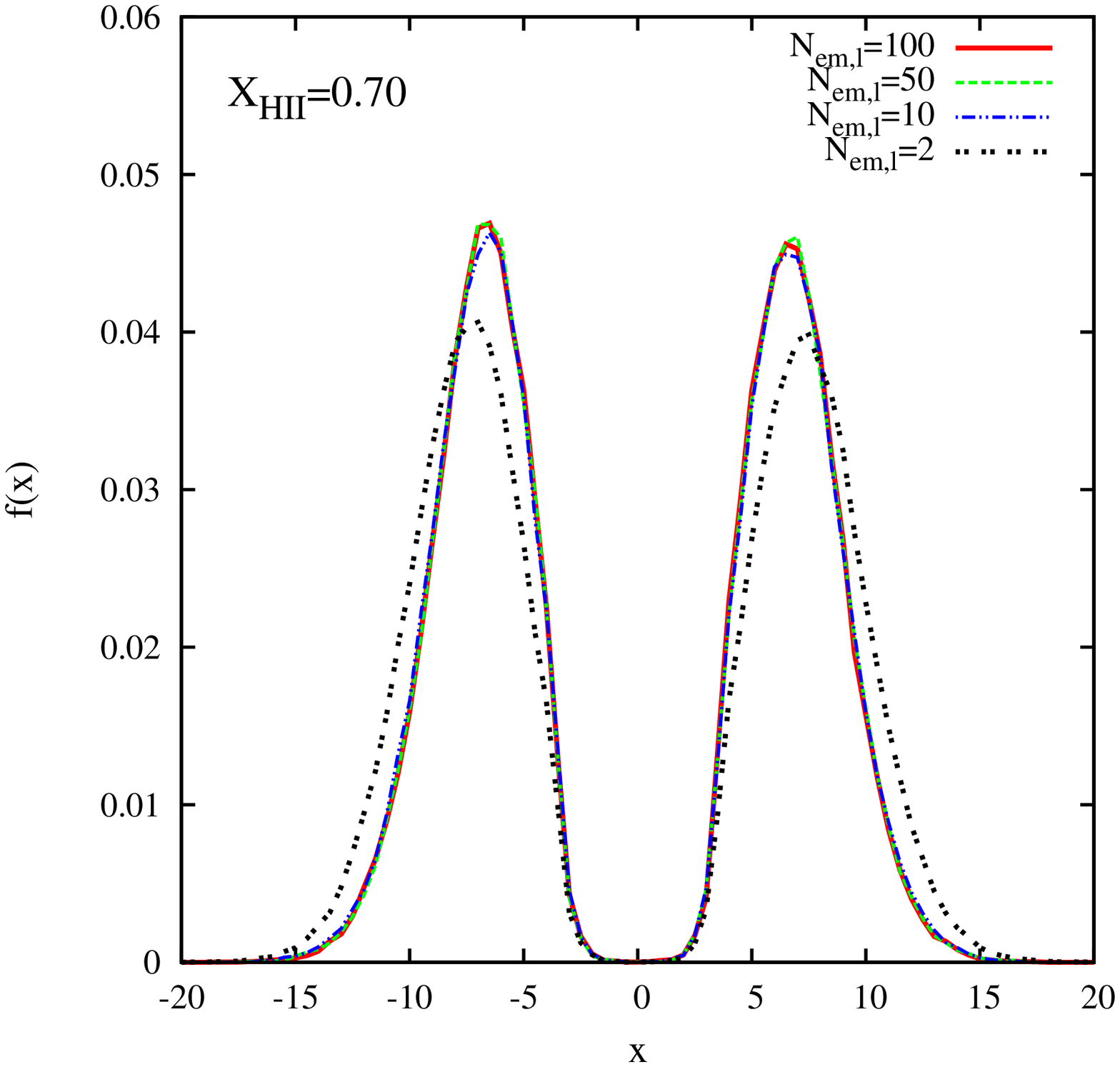}
   \includegraphics[width=7.5cm,angle=-90]{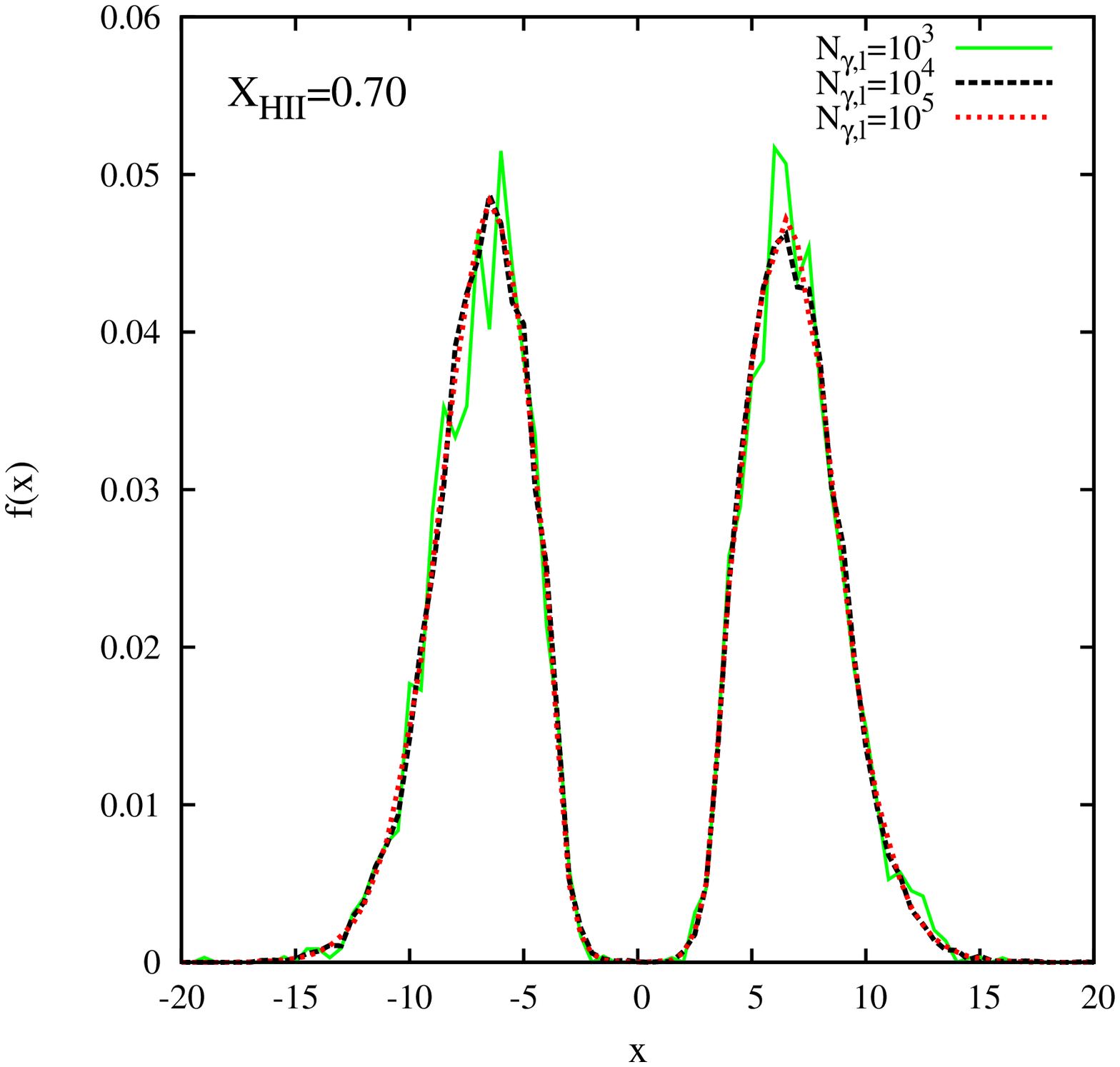}
  \caption{Effect of the input parameters on the final spectra.
The top-left plot shows \lya spectra integrated over the full simulation time ($N_{out}=1$)
using a different number of \lya emissions, $N_{em,l}$, and $N_{\gamma,l}=10^4$.
The top-right and bottom-left plots show the instantaneous spectrum at $\chi_{HII}=0.4$ and 0.7 respectively,
with the same values of $N_{em,l}$ and $N_{\gamma,l}$. Finally, the bottom-right plot shows a case
with $N_{em,l}=100$, $N_{out}=50$ and a spectrum at $\chi_{HII}=0.7$ with different values of $N_{\gamma,l}$.}
   \label{input_fig}

\end{figure*}

In this Section we investigate the dependence of the final spectra
on the parameters  $N_{em,l}$, $N_{\gamma,l}$ and $N_{out}$
defined in Sections \ref{lya_em} and \ref{spectrum}. The
convergence tests  presented are performed by adopting the same
conditions of the static case (see Sec.~4.1).

First we compare spectra built by integrating  over the full 
simulation time, {\it i.e.} we set $N_{out}=1$.
We have done several runs varying the number of
\lya emissions ($N_{em,l}= 2,10,50,100$) and setting $N_{\gamma,l}=10^4$.
\lya photons are emitted according to equation~\ref{em_2}, {\it i.e.} 
at regular ionization intervals.
For $N_{em,l}=2$, the two emissions are performed at the beginning and
once the ionization degree has become stationary.
The results are shown in the top-left panel of Figure~\ref{input_fig}.
From an inspection of the Figure it is clear that 2 emissions are
not accurate enough to properly describe the emergent \lya spectrum.
This is a consequence of the method described in Section 3.1 for weighting
the contribution to the spectrum from photons emitted at
different steps. In fact, as the time between the two emissions is large,
the weight assigned to the photons of the second (and last) emission
is so large that the contribution from the photons of the first emission
is negligible. Moreover, at the time of the second emission ionization
is almost complete and as a consequence the spectrum has two thin peaks
very close to the central frequency and shows no large frequency shift
due to scattering in a gas with larger opacity. So, in this case the outcoming spectrum
is the result of \lya transfer in an almost ionized sphere and the stages
through which the gas reached such configuration are completely neglected.
As the number of emissions is increased, the accuracy improves and
convergence is reached when $N_{em,l}>50$. 

We have then checked convergence of instantaneous (vs integrated)
spectra, setting $N_{out}=10$ and performing again four runs with
different numbers of \lya emissions ($N_{em,l}=2,10,50$ and $100$) and 
$N_{\gamma,l}=10^4$.
Figure~\ref{input_fig} displays two sets of spectra, each corresponding to a
fixed $\chi_{\rm HII}$ value: 0.4 (top-right panel) and  0.7 (bottom-left panel).
Effectively, the spectrum corresponding to $N_{em,l}=2$ is built only with
photons from the first emission, as the second is performed when $\chi_{\rm HII}>0.7$.
At $\chi_{\rm HII}=0.4$ the spectra are all very similar because
the gas opacity is sufficiently high to trap \lya photons and those
emitted close to the time when the spectrum is built are still scattering in the
gas. On the contrary, as ionization proceeds, photons emitted at later times encounter 
less neutral hydrogen and escape more easily, and their contribution to the outcoming
spectrum increases. As a consequence, the case with $N_{em,l}=2$, when \lya
photons have been emitted only at the beginning of the simulation, displays a broader
and more shifted profile.

Finally, we use different values of
$N_{\gamma,l}=10^3,10^4,10^5$, with $N_{em,l}=100$ and $N_{out}=50$, to determine
the smallest number of \lya photons needed in each emission
in order to achieve convergence.
In the bottom-right plot we show \lya profiles corresponding to
an ionization fraction of  $\chi_{\rm HII}= 0.7$. As expected, the increasing number of
\lya photons in every emission produces a smoother profile.
We find that, for the configuration considered, the emissions 
characterized by $10^3$ photons are not accurate enough, and at least
$10^4$ photons per emission are needed.

\section{Summary}
\label{summary}
In this paper we have presented \CRLya, the first radiative transfer code
for
cosmological application that follows the parallel propagation of \lya and
continuum photons.
Since \lya propagation is dominated by resonant scattering with neutral
gas, the effect of a rapid change in the degree of gas ionization can 
affect the features of the emerging \lya spectrum.
To investigate this issue, we have developed in the continuum radiative
transfer
code \CR (Ciardi et al. 2001; Maselli, Ferrara \& Ciardi 2003;
Maselli \& Ferrara 2005, Maselli, Ciardi \& Kanekar 2008) a new algorithm to follow
the propagation of \lya photons through a gas configuration while it is changed by ionizing
radiation.

In order to perform the implementation, it has been necessary to introduce
the time evolution for \lya propagation, a feature commonly neglected in
line radiative transfer codes. This is a crucial aspect because, due to the
resonant
scattering nature of \lya transfer in a neutral medium, \lya radiation can
remain trapped for a substantial fraction of the simulation lifetime
before being
able to propagate away from its emission site, while the propagation time
of the ionization front can be much shorter.
Another challenge of the implementation has been to reduce the computation
time for the \lya scattering. In fact, to correctly model the \lya
propagation every single scattering should be followed. As this would require prohibitively
large computational times, we have used a statistical approach to the \lya
treatment.
We have compiled tables using MCLy$\alpha$ (Verhamme, Schaerer \& Maselli
2006) to describe the physical characteristics of a photon
escaping from a gas cell where it was trapped by scattering
as a function of the temperature and density of the gas as well as of
the incoming photon frequency. The tables are called within \CRLya.
With this statistical approach we experience a drastic reduction of
the computational time and, at the same time, an excellent agreement with
the full \lya radiative transfer computations.

We have discussed the details of the code implementation
and tested it for several gas configurations, including 
static spherical gas distribution, expanding and collapsing spheres
and expanding shell. 
For all the configurations analyzed, \CRLya reproduces emerging
spectra with the qualitative features expected from theoretical models and
discussed previously in the literature, while a more quantitative
comparison has not been feasible as \CRLya is the first code which
couples the continuum and line transfer. 
With this implementation, it has also been possible to investigate how
the line shape of the emergent spectra evolves with the gas ionization.
Although the specific results depend on the geometry of the gas and on
the velocity field, common trends are found. 
The main results can be summarized as follows.

\begin{itemize}

\item While ionization proceeds the peaks on the blue/red side of the
line center move closer to the central frequency, getting thinner and higher, as
expected for a gas configuration with progressively decreasing
optical depth. Depending on the gas configuration
({\it e.g.} in case of an expanding shell), more complex features arise 
that can be associated to the different paths followed by the \lya photons before escaping.
\item The emerging spectra keep memory of the ionization history which
generates a given gas configuration.
\item The novel approach to \lya transfer developed in \CRLya allows
to resolve the emergence of different spectral features at different
times during the evolution of the ionization field. Features emerging
on different time scales are typically associated to the various paths
traveled by the photons before escaping. 
\item A comparison between our new algorithm to follow the propagation
of \lya photons and a full line radiative transfer shows an excellent 
agreement for different gas configurations and an enormous gain in 
computational speed.
\end{itemize}

In order to account for the effects discussed above a self-consistent
joint evolution of  line and ionizing continuum radiation as implemented
in \CRLya is necessary. A comparison between the results from our code and from
\lya scattering alone on a fixed density field shows that the extent of the difference between
the emerging spectra depends on the particular configuration considered, but it can be
substantial and can thus affect the physical interpretation of the problem at hand.

A detailed discussion on which are the objects/configurations for
which the ionization effects are expected to be relevant in shaping
the observed \lya spectrum is deferred to a forthcoming publication. 
Nevertheless, we have here discussed two specific test configurations for
which the coupling of continuum and line radiation would be necessary
in order to recover correctly the emergent profile. 
These differences are due to the time evolution feature introduced in
\CRLya for \lya photons which allows to keep track of the
ionization history imprint on \lya profiles for the first time in the literature.

The time evolution that builds up the \lya radiation field
can be furthermore important when calculating the impact of such
radiation on gas properties like the spin temperature, relevant
to predict the observability of 21~cm radiation from the early universe.  
In a forthcoming extension of the code \CRLya, we plan
to include the self-consistent calculation of the impact of the \lya 
radiation on the gas temperature, together with the contribution to
the \lya radiation from recombinations occurring in the gas.


\section*{Acknowledgments}
The authors thank Anne Verhamme and Daniel Schaerer for providing the
last version of the code MCLy$\alpha$ and for their comments. 
They are also thankful to Andrea Ferrara for stimulating
discussions and sharp comments on the draft. 
AM is supported by the DFG Priority Program 117.

\begin{appendix}

\section{Tables for Ly$\alpha$ scattering}

Following the single scatterings of each Ly$\alpha$ photon can be
extremely expensive; in order to avoid it we have built tables by using a statistical 
approach that allows to retrieve the frequency of the outcoming photon, $x_{out}$, and the 
time interval for which the photon is trapped in the gas by the scatterings, $t_{scatt}$, given 
the frequency of the incoming photon, $x_{in}$, the gas temperature, $T_{cell}$, 
and optical depth, $\tau_{cell}$, of the cell where the scattering takes place. 
We adopt the opacity at line center to characterize the optical depth of the gas in a cell.
Notice that the frequencies are always meant in the comoving frame.

The tables are compiled in the following way. Given a value for the input parameters ($T_{cell}$, 
$\tau_{cell}$ and $x_{in}$), we run the MCLy$\alpha$ code several times (the results
converge when 10000 \lya photons are emitted) to obtain values for $x_{out}$ and $t_{scatt}$
which are then binned in distribution functions.
The $x_{out}$ distribution is binned using regular intervals of 0.5.
For $t_{scatt}$ we adopt equally spaced logarithmic bins of 0.014. It is important to note that 
the value of $t_{scatt}$, which depends on the distance traveled by the photon, is linked
to the size of the cell. Thus, the tables are compiled for a reference cell size, $d_{c,ref}$,
but anytime they are accessed by \CRLya, the value of $t_{scatt}$ obtained needs to be rescaled
for the actual cell dimension, $d_c$.
The ranges that the tables cover are:
\begin{itemize}
 \item Temperature: $10 \; {\rm K} \leq T \leq 10^5 \; {\rm K}$
 \item Optical depth: $1  \leq \tau \leq 10^6 $
 \item Frequency: $-100 \leq x \leq 100$
\end{itemize}

\begin{figure}
\begin{center}
{\includegraphics [width=8cm,angle=-90] {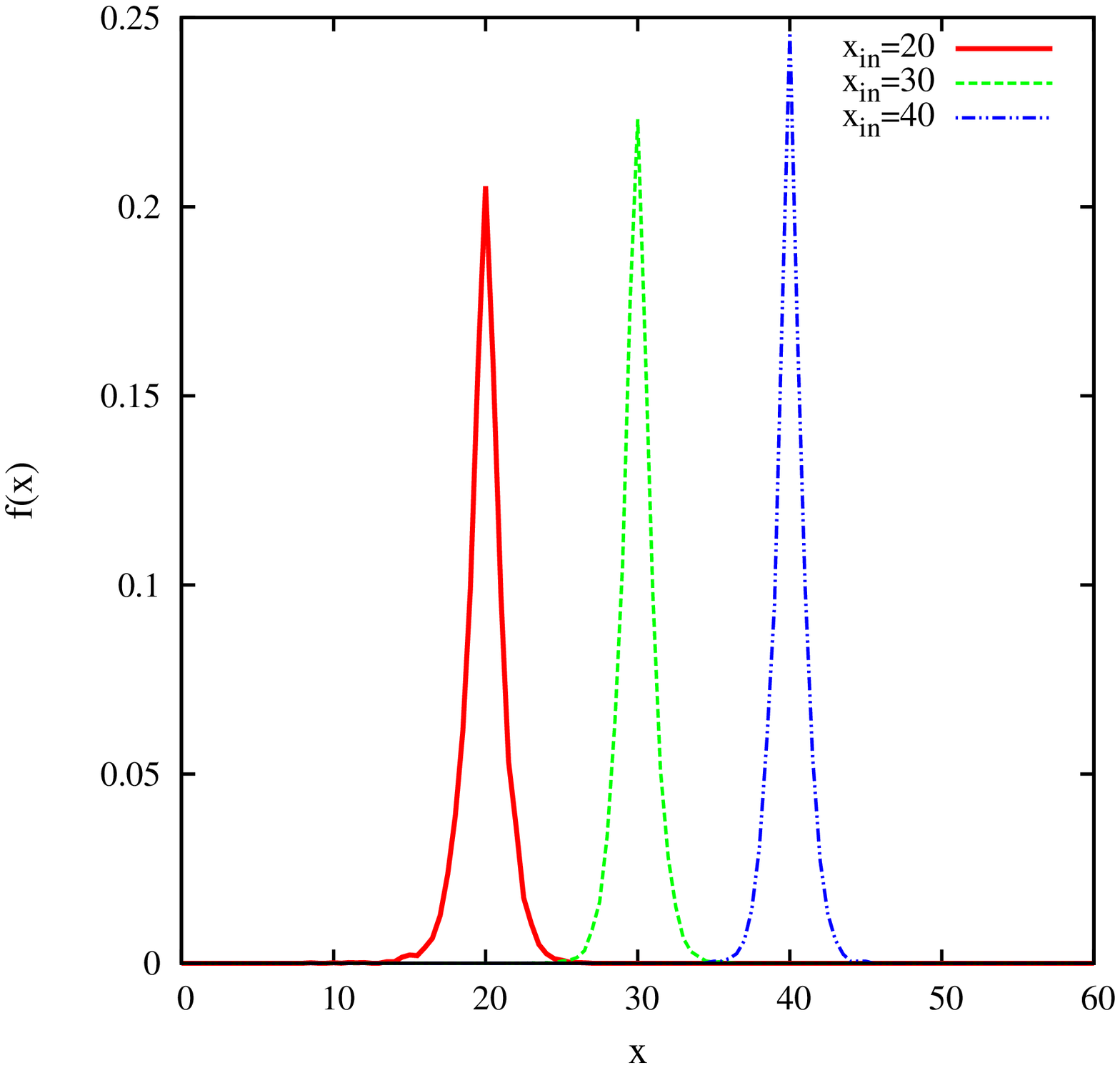}}
\caption{Distribution of photon outcoming frequencies generated by multiple scatterings inside a 
cubic cell with $T_{cell}=10^4$~K and $\tau_{cell}=10^6$ for an incoming photon frequency
of $x_{in}=20$ (solid line), 30 (dashed line) and 40 (dashed-dotted line).}
\label{x_out} 
\end{center}
\end{figure}

\begin{figure}
\begin{center}
{\includegraphics [width=8cm,angle=-90] {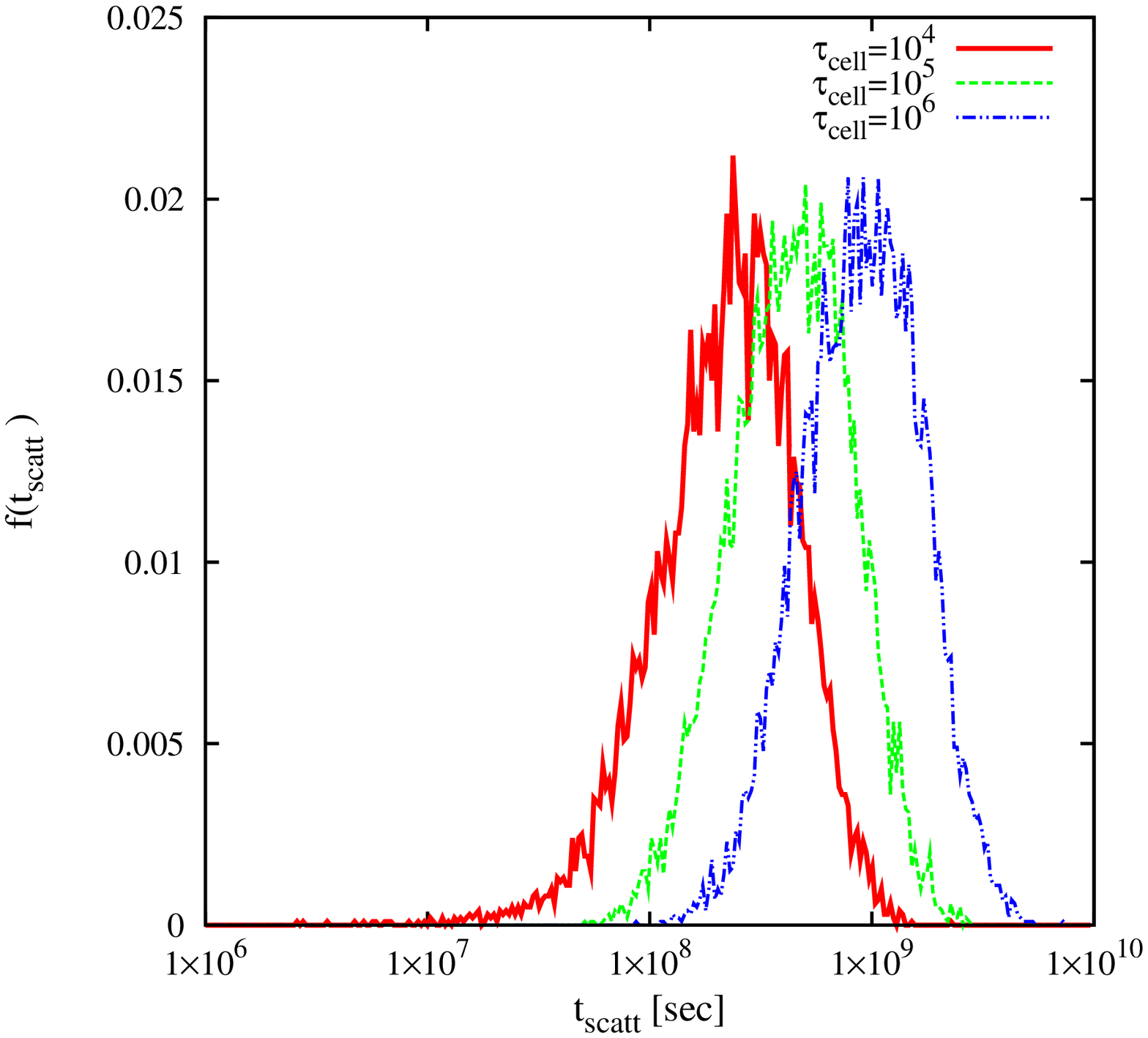}}
\caption{Distribution of the time that a photon with $x_{in}=0$ spends in a cubic cell with 
$T_{cell}=10$~K and an optical depth $\tau_{cell}=10^4$ (solid line), $10^5$ (dashed line)
and $10^6$ (dashed-dotted line). }
\label{t_cell} 
\end{center}
\end{figure}

An example of $x_{out}$ and $t_{scatt}$ distributions is given in Figures \ref{x_out} and \ref{t_cell}.
Figure \ref{x_out} shows the distribution of $x_{out}$ for $T_{cell}=10^4$~K, $\tau_{cell}=10^6$ and
different values of $x_{in}$ ($x_{in}=$20, 30 and 40).
Figure \ref{t_cell} shows how the  $t_{scatt}$ distribution changes for different values of the optical depth,
$\tau_{cell}=10^4,10^5,10^6$; here we fixed $T_{cell}=10$~K and $x_{in}=0$.

The above tables are accessed by \CRLya in the following way.
If $T_{cell}$, $\tau_{cell}$ and $x_{in}$ are within the range covered by the tables, 
the distributions for $t_{scatt}$ and $x_{out}$ are calculated by a linear interpolation
and then the value used in \CRLya for $t_{scatt}$ and $x_{out}$ is obtained by MC sampling the
interpolated distributions. 
The interpolation scheme for $t_{scatt}$ is as follows.                 
As a first step, the values of temperature $T_1$ and $T_2$ closest to $T_{cell}$ for
which $T_1\leq T_{cell}<T_2$ are found. The same is done for the optical depths
$\tau_1$ and $\tau_2$, and the frequencies $x_1$ and $x_2$.
The weights to be assigned to each value are derived by linear interpolation. 
As an example we can consider a simple 1D case, with linear interpolation only on
temperatures. In this case, $t_{scatt}=w_{T_1} t_{scatt}(T_1) + w_{T_2} t_{scatt}(T_2)$,
where the weights are $w_{T_1}=(T_2-T_{cell})/\Delta T$ and 
$w_{T_2}=(T_{cell}-T_1)/\Delta T$, with $\Delta T=T_2-T_1$. 
The same procedure, extrapolated in 3D, produces a distribution of $t_{scatt}$
that will be randomly sampled in \CRLya.
The interpolation for $x_{out}$ follows the same steps, after a shift in the frequency space has 
been performed to assure that the correct distribution is obtained and no spurious peak forms.        
To understand how the $x_{out}$ interpolation works an easy example can be useful.
Let us assume that our purpose is to reproduce the dashed profile centered in $x=30$ (Fig.~\ref{x_out})
starting from the two profiles at $x=20$ and $40$.
If the interpolation were performed bin by bin without any previous shift ({\it e.g.} the bin [19.5-20[
of the solid curve with the bin [39.5-40[ of the dashed-dotted curve, and similarly for all the bins),
the result would be two smaller peaks centered at $x=20$ and $40$, rather than one centered at $x=30$.
To perform a correct interpolation we need to center the solid and dashed-dotted profiles on $x=30$
and then interpolate. 

\begin{figure}
\begin{center}
{\includegraphics [width=7.5cm,angle=-90] {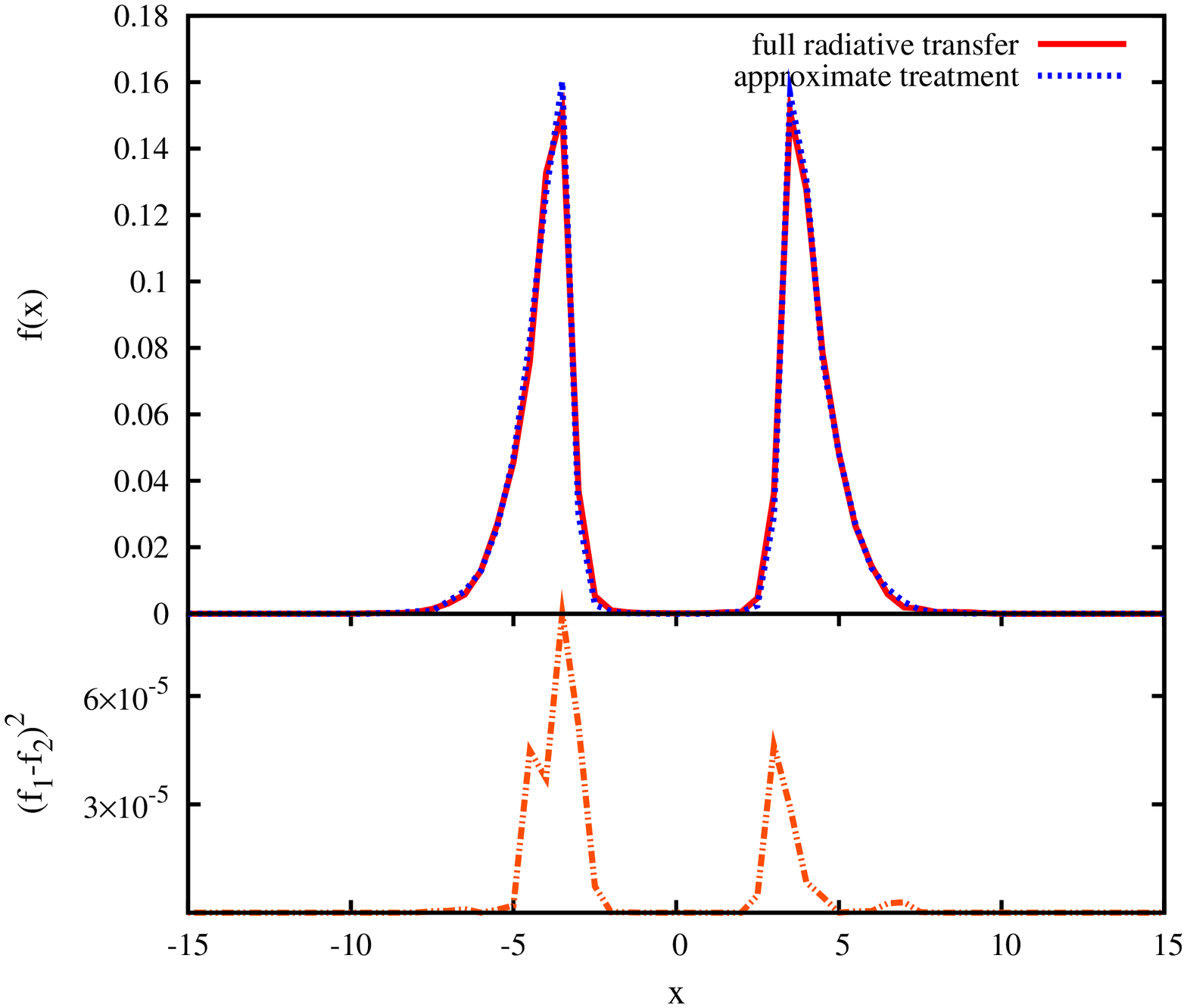}}
\caption{The upper panel shows the predicted Ly$\alpha$ frequency distribution for monochromatic
line radiation escaping from a gas with temperature $T=8000$~K and optical depth $\tau=10^6$.
The dotted (solid) line indicates the results for the approximate
(full radiative transfer) treatment using tables build with 10000 photons.
The difference between the two distributions is shown in the bottom panel.}
\label{comparison} 
\end{center}
\end{figure}

\begin{figure}
\begin{center}
{\includegraphics [width=7.5cm,angle=-90] {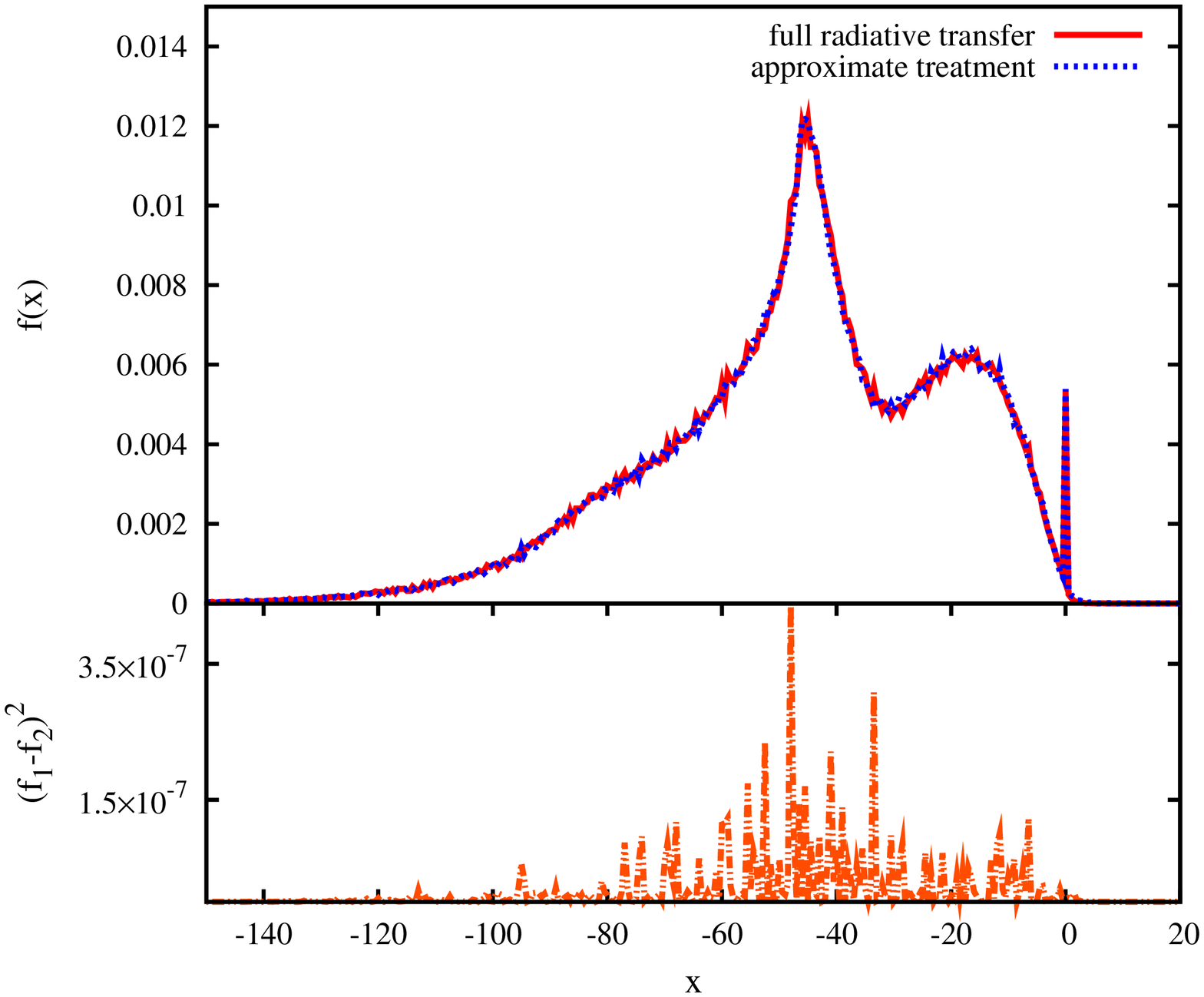}}
\caption{As Fig.~\ref{comparison} but for a gas configuration resembling a shell
with temperature $T=10^4$~K and column density $N_{\rm HI}=2 \times 10^{20}$~cm$^{-2}$
expanding with an uniform radial velocity $V=300$~km~s$^{-1}$.
The dotted (solid) line in the upper panel indicate the results for the approximate
(full radiative transfer) treatment using tables build with 400000 photons.}
\label{comparison_shell} 
\end{center}
\end{figure}

To check the validity of our approach, we have compared
the results from the full radiative transfer treatment (using MCLy$\alpha$) with a case in which the tables were used. In
Figure \ref{comparison} the distribution in frequency of 10000 Ly$\alpha$ photons escaping from a
gas with temperature
$T=8000$~K, optical depth $\tau=10^6$ and frequency $x=0$ is shown. The dotted (solid) line in the 
upper panel indicates the results for the approximate
(full radiative transfer) treatment using tables built with 10000 photons. 
In the bottom panel the difference between
the two distributions is plotted, showing an excellent agreement, which is found also for different initial
conditions. 
We perform the same check with a more complex gas distribution using the expanding shell 
described in Section~\ref{shell}. As the final spectrum is expected to have a wider frequency 
distribution, this test is performed to check the accuracy of the $x_{out}$ interpolation at
larger frequency shifts. 
The results are shown in Figure \ref{comparison_shell}; for both profiles we have used 400000 photons.
Also in this case an excellent agreement is found.

As the changes in the gas properties during a simulation can be drastic, sometimes the values
$\tau_{cell}$ and $x_{in}$ can fall outside the range covered by the tables (we do not
expect $T_{cell}$ to fall outside the range).
In these cases we perform an extrapolation of the existing tables. More in particular,
for $\Vert x_{in}\Vert>100$ we use the same distributions derived for $\Vert x_{in}\Vert=100$.
This is a good approximation as at these frequencies the cross section is small and \lya
scattering rare. The extrapolation for the optical depth works differently. 
If $\tau_{cell}<1$ no interaction takes place and the photon propagates freely. 
If $\tau_{cell}>10^6$ we divide the cell into $2^{3n}$ sub-cells (where $n$ is the number
of divisions performed) until each sub-cell has an optical depth $<10^6$. At this point,
every sub-cell is treated as a single cell.

The use of the tables allows an enormous gain in computational speed.

\end{appendix}

\label{lastpage}

\end{document}